\documentclass[11pt]{article}
\usepackage{graphicx,amsmath,amsfonts,amssymb}
\def\sloppy{\tolerance=100000\hfuzz=\maxdimen\vfuzz=\maxdimen}
\def \beq  {\begin{equation}}
\def \eeq  {\end{equation}}
\def \beqar {\begin{eqnarray}}
\def \eeqar {\end{eqnarray}}
\def\bsp{\beq\begin{split}}
\allowdisplaybreaks
\mathchardef\mhyphen="2D

\def\la {{\langle}}
\def\ra {{\rangle}}
\def\vx {{\vec x}}
\def\vy {{\vec y}}
\def\vk {{\vec k}}
\def\vf {{\varphi}}

\def\Tr {{\rm Tr}}

\def\bD {\bar{D}}

\def\by {\bar{y}}

\def\vk {\vec{k}}
\def\vx {{\vec x}}

\def\vy{\vec{y}}
\def\vv {\vec{v}}
\def\vu {\vec{u}}

\def\del {\partial}
\def\bdel{\bar{\partial}}

\def\b {\beta}
\def\e {\epsilon}
\def\d {\delta}

\def\l {\lambda}

\def\bz {{\bar{z}}}

\def\G {{\cal G}}

\def\vf {{\varphi}}

\def\half{\textstyle{1\over 2}}

\begin{document}
\def \CMP {{Commun. Math. Phys.}}
\def \PRL {{Phys. Rev. Lett.}}
\def \PL {{Phys. Lett.}}
\def \NPBProc {{Nucl. Phys. B (Proc. Suppl.)}}
\def \NP {{Nucl. Phys.}}
\def \RMP {{Rev. Mod. Phys.}}
\def \JGP {{J. Geom. Phys.}}
\def \CQG {{Class. Quant. Grav.}}
\def \MPL {{Mod. Phys. Lett.}}
\def \IJMP {{ Int. J. Mod. Phys.}}
\def \JHEP {{JHEP}}
\def \PR {{Phys. Rev.}}
\def \JMP {{J. Math. Phys.}}
\def \GRG{{Gen. Rel. Grav.}}
\begin{titlepage}
\null\vspace{-62pt} \pagestyle{empty}
\begin{center}
\rightline{CCNY-HEP-12/2}
\rightline{January 2012}
\vspace{1truein} {\Large\bfseries
Supersymmetry and Mass Gap in 2+1 Dimensions:}\\
\vskip .1in
{\Large\bfseries
A Gauge Invariant Hamiltonian Analysis}\\
\vspace{6pt}
\vskip .1in
{\Large \bfseries  ~}\\
\vskip .1in
{\Large\bfseries ~}\\
{\large\sc Abhishek Agarwal$^{a,b}$ and V.P. Nair$^a$}\\
\vskip .2in
{\itshape $^a$\,Physics Department\\
City College of the CUNY\\
New York, NY 10031}\\
\vskip .1in{\itshape $^b$\,Physical Review Letters\\
American Physical Society\\ 
Ridge, NY 11367}\\
\vskip .1in
\begin{tabular}{r l}
E-mail:
&{\fontfamily{cmtt}\fontsize{11pt}{15pt}\selectfont abhishek@aps.org}\\
&{\fontfamily{cmtt}\fontsize{11pt}{15pt}\selectfont vpn@sci.ccny.cuny.edu}
\end{tabular}

\fontfamily{cmr}\fontsize{11pt}{15pt}\selectfont
\vspace{.8in}

\centerline{\large\bf Abstract}
\end{center}
A Hamiltonian formulation of Yang-Mills-Chern-Simons theories with $0\leq \mathcal{N}\leq 4$ supersymmetry in terms of gauge-invariant variables is presented, generalizing earlier work
on nonsupersymmetric gauge theories.
Special attention is paid to the volume measure of integration (over the gauge orbit space of the fields) which occurs in the inner product for the wave functions and arguments relating it to
the renormalization of the Chern-Simons level number and to mass-gaps in the spectrum of the Hamiltonians are presented.
The expression for the integration measure is consistent with the absence of mass gap for
theories with extended supersymmetry (in the absence of additional matter hypermultiplets and/or Chern-Simons couplings),
while for the
minimally supersymmetric case, there is a mass-gap, the scale of which is set by a renormalized level number, in agreement with indications from existing literature.
The realization of the supersymmetry algebra and the Hamiltonian in terms of the gauge invariant variables is also presented.
\end{titlepage}
\pagestyle{plain} \setcounter{page}{2}
\section{Introduction and Summary}

In this paper we present a reformulation of $\mathcal{N} = 1,2$ and $4$ supersymmetric Yang-Mills theories (with and without Chern-Simons couplings) in three spacetime dimensions
in terms of gauge-invariant variables, following the Hamiltonian approach in \cite{KKN-long}. As part of this reformulation, we compute the volume measure on the physical gauge-invariant configuration space $\mathcal{C}$ of the theories under consideration and study the interplay between dynamical mass-generation and supersymmetry in $D=2+1$.

Studies of gauge theories in three dimensions, especially their nonperturbative aspects, are motivated by several current issues of both practical as well as conceptual importance. For example, it is well known that various equilibrium properties of the quark gluon plasma (QGP) are expected to be captured by an Euclidean three dimensional Yang-Mills theory coupled to scalar adjoint matter fields. The nonperturbative mass-gap of the three dimensional gauge theory, in this context, sets the screening length for chromomagnetic fields in the plasma.  Thus 
the QGP provides a very physically relevant context for gauge dynamics in three dimensions;
a detailed description relating non-perturbaitve effects of the three dimensional theory to physical observables of the plasma can be found in the rather large literature on the subject; see, for example \cite{qgp}.  On the more conceptual side, supersymmetric Yang-Mills and Chern-Simons theories appear at the forefront of several exciting recent developments pertaining to the gauge-gravity dualities involving $M2$ and $D2$ brane theories \cite{blg, abjm}. For instance, understanding  both the flow of the $D2$ brane theory to a conformal field theory in the IR
and its connection to a holographic description of high $T_c$ superconductivity \cite{hightc} require tools for probing the nonperturbative behavior of $\mathcal{N}=8$ SYM.

Keeping these broad motivations in mind it is interesting to note that,
in the case of the purely gluonic theory, it has been possible to do {\it ab-initio} strong coupling computations\footnote{The wording ``strong coupling computations" is meant to indicate a specific expansion scheme, as explained in detail in the second paper in \cite{KKN-string}.}
 of certain physical quantities using the gauge-invariant Hamiltonian formalism
\cite{KKN-long}.  For example, systematic computations of the string tension \cite{KKN-string}, screening effects \cite{scalar} and the inclusion of the effects of non-trivial spatial 
geometries \cite{sphere} have been studied within this framework. 
The Hamiltonian approach, with some variations, has also been employed to estimate the glueball spectrum in \cite{glueballs}. Further, the formalism has been extended \cite{KKN-CS} to the Yang-Mills-Chern-Simons theory \cite{DJT}.
Most importantly, at least as far as the focus of the present paper is concerned, the origin of the mass-gap in the purely gluonic theory can be understood in a geometric fashion as the effect of the volume measure on $\mathcal{C}$, a fact made transparent 
in terms of manifestly gauge-invariant variables
\cite{KKN-long, robustness}.  

On the other hand, a very different set of tools allows one to make considerable headway into the study of gauge theories in $D=2+1$, as long as one has supersymmetry.  For the case of Yang-Mills and Chern-Simons theories with extended supersymmetries many powerful statements can be made about their partition functions \cite{Z}, S-matrices \cite{so(8), scs-s}, physical spectra \cite{lm, spectrum} and Wilson-loop expectation values \cite{wilson-loops}. Obviously, in the case of maximal $(\mathcal{N}=8)$ or near-maximal $(\mathcal{N}=6)$ supersymmetry one also has a web of gauge-gravity dualities which can be used to make precise statements about the strong coupling behavior of $\mathcal{N}=8$ Yang-Mills and $\mathcal{N}=6$ Chern-Simons theories\cite{lm, abjm}. The potent methods that enable these computations often use manifest supersymmetry in one way or another and are not readily generalizable to non-supersymmetric theories.  A natural question to ask is if these two approaches, namely, the gauge-invariant Hamiltonian point of view due to KKN \cite{KKN-long}, which does not rely on supersymmetry in any way, and the powerful computational tools that have been so successful in the recent studies of  supersymmetric gauge theories in three dimensions, can be fruitfully combined. In this paper we take a step towards answering this question. A summary of our main results and the outline of the organization of the paper are as follows.

Our particular focus in this paper will be on Yang-Mills-Chern-Simons theories with 
$0\leq \mathcal{N} \leq 4$ supersymmetry with particular emphasis on the issue of dynamical mass-generation in the presence of supersymmetry. We re-express these theories in a gauge-invariant  form in a Hamiltonian framework by extending the KKN formalism to incorporate supersymmetry. In the process, we also compute the integration measure for the inner product of the wave functions for these theories. As discussed in the next section, in many ways, this is the quantity of central interest
as far as the existence of a mass gap is concerned.
For the pure glue theory, this measure is given in terms of a Wess-Zumino-Witten (WZW) functional (\ref{vol3}) with a level number equal to $2\, c_A$, where $c_A$ is the quadratic Casimir  value for the adjoint representation of the gauge group (which we shall take as $SU(N)$)  \cite{GK, Bos-Nair}. Ultimately, this level number 
determines the mass-gap in the spectrum of the theory \cite{KKN-long}, the basic argument for which is also outlined in this section.
We also discuss the effect of the Chern-Simons coupling (when it is present), and its renormalization, on the measure and the mass-gap.
Much of the discussion in this section is based on previous work \cite{KKN-long, KKN-CS, Bos-Nair};
 it is intended to serve as a coherent summary of the interrelations among the three key quantities:
 the integration measure for the inner product of wave functions (a quantity defined within a Hamiltonian approach),
the mass-gap, and the renormalization of the Chern-Simons level number (which is carried out in a covariant functional integral approach).

As is well known, in the absence of supersymmetry, the level number for the Chern-Simons theory, $k$, with or without the presence of a Yang-Mills term in the Lagrangian, does, indeed, undergo a renormalization $k \rightarrow k + c_A$ \cite{pisarski-rao}.
(This can be seen from a Hamiltonian point of view as well, see \cite{witten-top, Bos-Nair}.)
For the Yang-Mills-Chern-Simons system, this  contribution can also be computed
via Feynman diagrams in one-loop covariant perturbation theory \cite{pisarski-rao}; the supersymmetric
extension, for $1\leq \mathcal{N}\leq 4$ Yang-Mills-Chern-Simons theories, 
has been done in \cite{lee-scs}
Our arguments show how the
coefficient of this shift, $c_A$, is, in fact, directly related to the level number of the
 WZW functional (\ref{vol2}) occurring in the inner product for the wave functions.
This nonperturbative origin of this shift also clarifies why the same shift is obtained for both the Yang-Mills-Chern-Simons system \cite{pisarski-rao, KKN-CS} as well as the pure Chern-Simons theory \cite{witten-top, Bos-Nair}.

In preparation for the study of supersymmetric theories, section \ref{fermi-measure} is devoted to  a more direct computation of the contribution of Majorana fermions to the measure, when the fields are recast in gauge-invariant forms. An important point is that there are different ways to define gauge-invariant variables for the fermions, with correspondingly different Jacobians and
integration measures.  The calculation of the Jacobian
is related to a chiral anomaly calculation. The choice of the gauge-invariant version of the fermions is related to the realization of the supersymmetry algebra; this is explored in the later sections.

The gauge-invariant reformulation of the supersymmetric theories is taken up from
the next section onwards, with ${\mathcal N} = 1$ in section 4,
${\mathcal N}\geq 2$ in section 5, linearization in terms of the gauge-invariant variables in section 6, and some useful formulae collected in an appendix.
We construct the supercharges and their algebra, which then determine
the choice of gauge-invariant variables 
 and the fermionic contribution to the measure.
 The effect of the fermions is to eliminate
 the WZW functional
for extended supersymmetry ($\mathcal{N} \geq 2$), 
 while canceling the gluonic contribution by a factor of half in the minimal supersymmetric case
 (${\mathcal{N} = 1}$); i.e., $k \rightarrow k+c_A - {\half} c_A = k + {\half} c_A$.
 
 The expression for the measure indicates that one cannot have spontaneous mass-generation in supersymmetric Yang-Mills theories in three spacetime dimensions in the presence of $\mathcal{N} \geq 2$ supersymmetry, at least in the absence of additional matter hypermultiplets and as long as supersymmetry remains unbroken.  This statement pertains to theories that do not have Chern-Simons couplings, the presence of which make corresponding gauge theories massive by construction. The absence of spontaneous mass generation in the $k\rightarrow 0$ limit corresponds in our set up to a non-renormalization of the Chern-Simons level number for non-zero values of $k$.
 In the case of minimal supersymmetry, one cannot consistently define a Yang-Mills theory without the Chern-Simons terms due to the parity anomaly\cite{witten-index, n=1-lattice}. In this case,
 our results are tantamount to the calculation
 of the shift of the Chern-Simons level number - which sets the scale for the massive excitations of the gauge theory - as part of the computation of the integration measure exactly.

The implications of these results are also consistent with several other indications obtained in the previous literature. In particular, the
massless nature of the physical spectrum of Yang-Mills theories with $\mathcal{N} = 2$ and $4$ supersymmetry  is also expected based on D-brane constructions \cite{N=2-brane}, small volume arguments \cite{unsal-bion} and (specifically for the case of $\mathcal{N}\geq 4$) arguments based on the moduli space of the Coulomb branch \cite{seiberg-witten}.

Not surprisingly, the expressions for the supercharges and the Hamiltonian are nonlocal in terms of the gauge-invariant variables.
Nevertheless, we can verify the
required commutator algebra; 
the details are presented for the $\mathcal{N}=1$ theory.
There are also some regularization issues involving expressions like
Green's functions at coincident points; some details of how this can be done in a way compatible with the measure calculation are also given in the relevant sections. Several useful formulae, including regularized expressions for the Green's functions,  can be found in the appendix at the end of the paper.

Before proceeding to the  detailed analyses, let us remark that there are no natural obstructions to extending these present methods to theories with higher supersymmetry and to theories with fundamental matter fields.  We intend to explore the application of the formalism presented in this paper to  to D2 and M2 worldvolume theories and their deformations involving fundamental matter fields elsewhere. 
\section{The measure of integration on ${\cal A}/ {\cal G}_*$}

The geometry of the configuration space and the volume of integration for the wave functions have a crucial bearing on the question of a mass gap in the theory. It is useful to start with a general discussion of this issue for YM(2+1), the ${\cal N} =0$ theory, and then generalize from there.

\subsection{Yang-Mills theory in 2+1 dimensions}

We will start with some notational preliminaries. We will take the group in which the gauge transformations take values, often called the gauge group, to be $SU(N)$.
We will denote by $\{ t^a\}$, $a= 1, 2, \cdots, N^2-1$,
a  set of linearly independent traceless hermitian $N\times N$ matrices which form a basis for the
Lie algebra of $SU(N)$ in its fundamental representation. These matrices will be normalized by the condition $\Tr (t^a t^b ) = \half \delta^{ab}$ and obey the commutation rules
$[t^a , t^b ] = i f^{abc} t^c$. The structure constants $f^{abc}$ also define the adjoint representation of 
$SU(N)$ via $(T^a)_{bc} = -i f^{abc}$.

The gauge potential which is a vector field in 
2+1 dimensions is given by the matrix  $A_\mu =  A^a_\mu (-it^a)$, with the gauge
transformation given by
\beq
A_\mu \rightarrow A^g_\mu = g^{-1} A_\mu g + g^{-1} \del_\mu g 
\label{YM1}
\eeq
The Yang-Mills action, apart from terms involving superpartners (to be discussed later),  is given by
\beq
S = {1\over 2} \int d^3x~ \Tr (F_{\mu\nu} F^{\mu\nu} ) = -{1\over 4} \int d^3x~ F^a_{\mu\nu}F^{a\mu\nu} \label{YM2}
\eeq
where the
field strength tensor is given by $F_{\mu\nu} = (-it^a)F^a_{\mu\nu}
= \del_\mu A_\nu - \del_\nu A_\mu + [A_\mu , A_\nu ] = (-it^a) ( \del_\mu A^a_\nu - \del_\nu A^a_\mu
+f^{abc} A^b_\mu A^c_\nu )$.

In a Hamiltonian analysis, we can partially fix the freedom of gauge transformations
by setting $A_0 =0$. With this choice, the fields are $A_1$, $A_2$, the spatial components of the gauge potential. The wave functions are then functionals of $A_i$.
The Hamiltonian is given by
\beq
{\cal H} = \int d^2x~\left[ {e^2 E^a_i E^a_i \over 2}+ {B^a B^a \over 2 e^2}\right]
\label{YM3}
\eeq
where $E^a_i$ and $B^a$ are the electric and magnetic
components of the field strength, respectively, given by
\beq
E^a_i = \del_0 A^a_i = -i {\delta \over \delta A^a_i}, \hskip .2in
B^a = {1\over 2} \epsilon^{ij} \left( \del_i A^a_j - \del_j A^a_i + f^{abc} A^b_i A^c_j \right)
\label{YM4}
\eeq

The wave functions are functionals of the fields $A_i$. The choice of $A_0 =0$ still allows for gauge transformations by $g$ which are independent of time.  This implies a constraint on the wave functions, which is also the Gauss law or equation of motion for the $A_0$-component,
\beq
(\nabla _i \delta^{ab} + f^{abc} A^b_i ) E^a_i ~\Psi
= 0
\label{YM5}
\eeq
Since $E^a_i$ is the functional derivative with respect to $A^a_i$, this
is equivalent to requiring the invariance of the wave functions under gauge transformations
$g$ which go to the identity at spatial infinity. These are the true gauge transformations of the theory. In fact, it is useful to define
\beq
{\cal G}_* = \left\lbrace {\rm Set ~of ~all}~ g(\vx ) : {\mathbb R}^2 \rightarrow SU(N)~{\rm such ~
that~} g(\vx ) \rightarrow 1 {\rm as}~ \vert \vx \vert \rightarrow \infty \right\rbrace
\label{YM6}
\eeq
This has an action on the space of gauge potentials ${\cal A}$ as given in (\ref{YM1}),
where
\beq
{\cal A} = \left\lbrace {\rm Set ~of ~all ~Lie\mhyphen algebra\mhyphen valued ~vector ~functions}~A_i~{\rm on}~
{\mathbb R}^2 \right\rbrace
 \label{YM7}
\eeq
The space of physical (or gauge-invariant) configurations is then given by
${\cal C} = {\cal A} / {\cal G}_*$.
The statement that the wave functions are gauge-invariant, or equivalently, equation
(\ref{YM5}), can now be restated as saying that the wave functions are complex-valued functions (or more
generally, sections of a line bundle) on ${\cal C}$.

As a first step in relating the volume measure on ${\cal C}$ to the mass gap, notice that,
using
(\ref{YM3}) and (\ref{YM4}), the expectation value of the Hamiltonian for a state characterized by the wave function $\Psi$ is 
\beq
\la {\cal H}\ra = \int d\mu ({\cal C})~ \left[ {e^2 \over 2} {\delta \Psi^* \over \delta A^a_i (x)}
{\delta \Psi \over \delta A^a_i (x)} + {1\over 2 e^2} B^a(x) B^a(x)  \Psi^* \Psi \right]
\label{YM9}
\eeq
where $d \mu ({\cal C})$ denotes the volume element on ${\cal C}$.
Notice that the first term in $\la {\cal H}\ra$, arising from the kinetic energy term in (\ref{YM3}) can be
viewed as the gradient energy for $\Psi$ taken as a function on
${\cal C}$. This suggests that if the gauge-invariant distance between configurations 
cannot be made arbitrarily large, then the gradient energy cannot become arbitrarily small.
Such a geometric property of ${\cal C}$ can thus lead to a gap in the spectrum of
${\cal H}$. An argument for the mass gap along these lines, but at a very qualitative level, was advanced by Feynman in 1981 \cite{feynman}.
He suggested that while there are configurations (points)
whose separation in ${\cal A}$ is arbitrarily large, it is finite in  ${\cal C}$, after the equivalence under gauge transformations is taken into account.
This reasoning is indeed on the right track, but the desired property, as stated, does not hold
\cite{config1}.
We first show this and then see how the essence of the reasoning can be salvaged once the volume measure for ${\cal C}$ is taken into account.

The action for the theory can be written as
\beq
S = {1\over 2}  \int dt \int d^2x~ {\del A^a_i \over \del t}  {\del A^a_i \over \del t} 
~-~ {\rm potential ~terms}
\label{YM11}
\eeq
Comparing this with the action for a point particle,
\beq
S = {1\over 2} \int dt~ g_{ij} {d x^i \over dt} {dx^j \over dt} ~-~ {\rm potential ~terms},
\label{YM12}
\eeq
we see that (\ref{YM11}) can be interpreted as the action for a point particle moving on an infinite dimensional space, {\it viz.}, the space of fields, with metric
\beq
ds_{\cal A}^2 = \int d^2x~ \delta A^a_i \delta A^a_i
\label{YM13}
\eeq
The integrated version of this gives the
distance $s(A, A')$ between two gauge potentials $A$ and $A'$ as the
Euclidean distance, 
\beq
s^2 (A, A') = \int d^2x~ (A- A')^a (A-A')^a
\label{YM10}
\eeq
This metric  is still on the space of the gauge potentials. It may be possible to find a shorter distance between configurationss by choosing a different but gauge-equivalent potential for $A$ or $A'$. Thus, as the metric on the physical configuration space, we take \cite{config1, config2}
\beq
s^2_{\cal C} (A, A') = {\rm Inf}_g \int d^2x~ (A- A'^g )^a (A- A'^g)^a
\label{YM14}
\eeq

\subsection{The spikes}

With the definition (\ref{YM14}) of the distance on ${\cal C}$, we can now show that there are indeed configurations for which the separation can be arbitrarily large. It is enough to give example configurations to prove this point.
For this purpose, we may take one of the configurations as the vacuum, say, $A' =0$.
Then
\beq
s^2_{\cal C} (A, 0) = {\rm Inf}_g \int d^2x~ (A_i - g^{-1} \del_i g )^a (A_i - g^{-1} \del_i g )^a
\label{YM15}
\eeq
We will also use $SU(2)$-gauge theory, taking, as our example, the configuration
\beq
A = (-it^3) ~in (z\bz )^{n-1} {(z d\bz - \bz dz) \over [1+(z\bz )^n]}
\label{YM16}
\eeq
where $z = x_1 - i x_2$, $\bz = x_1 + i x_2$.
The minimum distance of the configuration (\ref{YM16}) from $A = 0$ can then be worked out as follows.
We parametrize the group element $g$ as
\beq
g = {1\over \sqrt{1+ f {\bar f}}} \left[ \begin{matrix} 
1&f \\
-{\bar f} &1\\
\end{matrix}\right] 
~\left( \begin{matrix} 
e^{-i\vf /2} &0\\
0&e^{i\vf/2}\\
\end{matrix}
\right)
\label{dist1}
\eeq
This leads to
\beq
s^2_{\cal C} (A,0) = {\rm Inf}_g \int d^2x~ \left[
\left( A - {i (f d{\bar f} - {\bar f} df) \over 1+f {\bar f} } + d\vf \right)^2 +
4 {\del_i {\bar f} \del_i f \over (1+ f {\bar f})^2} \right]
\label{dist2}
\eeq
Extremization with respect to $\vf$ is achieved if we choose 
$A - {i (f d{\bar f} - {\bar f} df) / (1+f {\bar f}) } $ to be transverse.
The first part of this expression, namely $A$, is already transverse and we can choose $f$ such that the second part is also transverse. The function $\vf$ which extremizes the
first term in (\ref{dist2}) is then $\vf =0$. Next we note that the expression for $s^2$ in (\ref{dist2}) consists of
two positive integrals. The minimum for the second term on the right hand side is given by
$8\pi Q[f]$, where $Q[f]$, which is an integer, is the topological charge of $f$, given by
\beq
Q[f] = {i\over 2\pi} \int d^2x~ \epsilon^{ij} { \del_i {\bar f} \del_j f \over (1+ f {\bar f})^2}
\label{dist3}
\eeq
The first term on the right hand side of (\ref{dist2}) is minimized, with a minimum value equal to zero, if we choose
$f = z^n$. In this case $s^2_{\cal C}(A,0)$ is $8\pi n$. Any other choice of
$f$ will lead to a larger value, because of the logarithmic divergence of the first term.
Thus we have shown that the distance of the configuration (\ref{YM16}) from the configuration 
$A'=0$, minimized with respect
to $g$, is given by
\beq
s^2_{\cal C} (A, 0) = 8 \pi n
\label{YM17}
\eeq
This tells us that, for any value of $L^2$, we can find configurations for which their separation from $A=0$ exceeds
$L$. The configuration (\ref{YM16}) is an example of these when
$n \geq (L^2 /8\pi )$.
The field strength corresponding to (\ref{YM16}) is
\beq
F = (-it^3) ~ (-4\, n^2) {(z\bz )^{n-1}  \over [1+ ( z\bz )^n]^2} ~dx_1\wedge dx_2
\label{YM18}
\eeq
Notice that $F$ is well-behaved, there is nothing pathological about it.

This concludes our argument that there are configurations, e.g., $A =0$ and
$A$ given by (\ref{YM16}) with $n \geq (L^2 /8\pi )$,  for which the minimal distance
between them, calculated in a gauge-invariant way, or equivalently on
${\cal C}$, can be arbitrarily large.
These are the so-called ``spikes" on ${\cal C}$.
Since the set of all physical configurations ${\cal C}$ is a connected space, there is obviously a line of configurations connecting, say,  $A=0$ to the $A$'s in (\ref{YM16}); more generally, a narrow spike running off from any configuration to infinity.

The spikes vitiate the simple argument for the gap by bounding the gradient energy
in (\ref{YM9}).
For, a wave function which is like a standing wave along this line would have arbitrarily long wavelengths suggesting that the kinetic energy
can be made infinitesimally small. However, this is not an adequate counter-argument for the gap, since we are in a multidimensional space and the measure of transverse directions is very important. A classic example of the importance of the transverse measure is given by
the two-dimensional Schr\"odinger problem
\beq
{\cal H} = - {\nabla^2 \over 2M} ~+~ \lambda (x^2 + x^2 y^2 )
\label{YM19}
\eeq
Notice that there is one direction, the $y$-axis ($x =0$), along which the potential energy is zero.
So one might expect long wavelength excitations along this direction, leading to a spectrum with no gap, connecting continuously to zero. But this is evidently too naive.
Such long wavelength excitations also have transverse oscillations, along the $x$-direction for which the
potential is $\lambda (1+ y^2 ) x^2$, corresponding to a frequency $\omega = \sqrt{ \lambda
( 1+ y^2)}$. The zero-point energy (or ground state energy) for this oscillation is $\half \sqrt{\lambda (1+y^2)}$.
This additional energy, increasing with $\vert y\vert$, cuts off the wave functions along the $y$-axis, vitiating the previous argument for the absence of a gap in the spectrum.
(Of course, in the full solution of the problem, it is not meaningful to separate the dynamics 
along $x$- and $y$-directions, as we do here. Nevertheless, our argument captures the essential physics.)

\subsection{The volume measure on ${\cal C}$}\label{volume-measure}

The question of whether the reasoning given for the Schr\"odinger problem
(\ref{YM19}) can hold for the Yang-Mills theory
in (2+1) dimensions hinges
clearly on the measure for the transverse directions. This has been calculated before and is given as follows.
The gauge potentials $A_z = ( {\half} (A_1 + i A_2 )$, $A_\bz = ( {\half} (A_1 - i A_2 )$ can be parametrized as
\beq
A_z = - \del_z M \, M^{-1}, \hskip .2in A_\bz = M^{\dagger -1} \del_\bz M^\dagger
\label{vol1}
\eeq
where $M \in SL(N, {\mathbb C})$ is a complex matrix. The hermitian matrix
$H = M^\dagger M$ is gauge-invariant; it parametrizes 
$SL(N, {\mathbb{C}})/SU(N)$.
The volume element for ${\cal C}$ is given by
\beq
d \mu ( {\cal C})  = d\mu (H) \exp [\,2~c_A~S_{wzw}(H)]
\label{vol2}
\eeq
where $S_{wzw}$ is the Wess-Zumino-Witten action given by
\beq
S_{wzw} (H) = {1 \over {2 \pi}} \int \Tr (\partial H ~\bdel
H^{-1}) +{i \over {12 \pi}} \int \epsilon ^{\mu \nu \alpha} \Tr (
H^{-1}
\partial _{\mu} H~ H^{-1}
\partial _{\nu}H ~H^{-1} \partial _{\alpha}H)
\label{vol3}
\eeq
In equation (\ref{vol2}), $d\mu (H)$ is the Haar measure for 
$H$ and $c_A$ is the value of the quadratic Casimir operator for the adjoint representation;
it is equal to $N$ for $SU(N)$. The inner product for wave functions is given by
$\la 1\vert 2\ra = \int d\mu ({\cal C})\, \Psi^*_1 \, \Psi_2$.
The total volume of ${\cal C}$ given by $\int d \mu ({\cal C})$ is the partition function
for the WZW model on $SL(N, {\mathbb{C}})/SU(N)$. It can be taken to be finite, with a suitable
regularization. The contrast to be made here is between the Abelian theory and the nonabelian one. For the former, since $c_A = 0$, the exponent in (\ref{vol2}) is zero and the integral
$\int d \mu ({\cal C})$ diverges for each mode.
For the nonabelian case, the integral is finite for each mode, although one does need a cutoff on the number of modes for the final result to be finite.
The ``finiteness" of $d \mu ({\cal C})$ suggests that the zero-point energy for the transverse dimensions should work to generate a mass gap for the theory
along the lines of the argument for (\ref{YM19}).

The fact that the measure is the key to the mass gap can be further elucidated in a couple of different ways. At a qualitative level, we can write, using the uncertainty principle,
\beq
\la {\cal H}\ra = {1\over 2} \left[e^2 \Delta E^2 + {\Delta B^2 \over e^2}\right]
= {1\over 2} \left[{e^2 p^2 \over \Delta B^2}+ {\Delta B^2 \over e^2}\right]
\label{vol4}
\eeq
where we consider modes of $E$, $B$ fields corresponding to a momentum value $p$. Ordinarily, to find the low lying modes,  we would minimize $\la {\cal H}\ra$ with respect to $\Delta B^2$ (obtaining $\Delta B^2 \sim e^2 p$)
to find $\la {\cal H}\ra \sim p$. (This would correspond to the photon in the Abelian theory.) 
However, in our case, the measure (\ref{vol2}) controls the dispersion in $B$ because, for low values of $p$, it becomes a very narrow Gaussian, since
\beq
S_{wzw} \approx  \left[ -{c_A\over 2\pi} \int B {1\over {p^2}} B +...\right]
\label{vol5}
\eeq
The resulting value of $\Delta B^2 = \pi p /c_A$ leads to
$\la {\cal H}\ra = (e^2 c_A /2\pi) + {\cal O} (p^2)$. We see the emergence of a mass gap from the properties of the measure.

At a more quantitive level, we can write the Hamiltonian in terms of the current
$J = (c_A /\pi) \del H \, H^{-1}$ as ${\cal H} = T +V$, with
\beq
{\cal H}= m \left[ \int_u J^a(\vu) {\d \over \d J^a(\vu)} ~+~ \int \Omega_{ab} (\vu,\vv) 
{\d \over \d J^a(\vu) }{\d \over \d J^b(\vv) }\right]
~+~ { \pi \over {m c_A}} \int \bdel J_a (\vx) \,\bdel J_a (\vx)
\label{vol6}
\eeq
where
\beq
\Omega_{ab}(\vu,\vv)= {c_A\over \pi^2} {\d_{ab} \over (u-v)^2} ~-~ 
i\, {f_{abc} J^c (\vv)\over {\pi (u-v)}} \label{vol7}
\eeq
The first term in (\ref{vol6}) assigns an energy of $m$ for each power of $J$.
This is the essence of the mass gap. The existence of this term is closely related to the
measure of integration. This term is needed to make ${\cal H}$ self-adjoint
for square integrable functions with the measure (\ref{vol2}).
To bring this out even more explicitly, we note that we
 can absorb the factor of 
$\exp ({ 2\, c_A \, S_{wzw}(H)})$ from the measure into the wave function by writing
$\Psi = \exp( - c_A S_{wzw}(H))\, \Phi$, the latter having the inner product
\beq
\la 1 \vert 2\ra = \int d \mu (H) \, \Phi_1^* \, \Phi_2
\label{vol8}
\eeq
The Hamiltonian acting on the $\Phi$'s is obtained by
\beq
{\cal H} \rightarrow {\cal H}_\Phi = \exp( c_A S_{wzw}(H)) \, {\cal H} \, \exp( - c_A S_{wzw}(H))
\label{vol9}
\eeq
If we use an expansion for $H$ as $H = \exp (\vf) \approx 1 + \vf +\cdots$,
$J = (c_A /\pi) \del \vf + \cdots$, then we can easily verify that
\beqar
\la 1 \vert 2\ra &\approx& \int  [d \vf ] ~~\Phi_1^* (H) \Phi_2
(H)\nonumber\\
{\cal H}_\Phi &\simeq& {1\over 2} \int_x \left[- {\d ^2 \over {\d \phi _a
^2 (\vx)}} + \phi _a (\vx)  
\bigl( m^2 -
\nabla ^2 \bigr)  \phi _a (\vx)\right] + \cdots
\label{vol10}
\eeqar
where $\phi _a (\vk) = \sqrt {{c_A k \bar{k} }/ (2 \pi m)}~
\vf _a (\vk)$. This clearly shows the emergence of the mass term.

\subsection{An alternate method for the measure of integration}

The arguments given above show the central role of the measure of integration in the inner product for the question of the mass gap.
We will now give an alternate way of obtaining this measure, which is more easily generalizable to the supersymmetric case.
It will rely on relating the measure to the Chern-Simons theory and the Yang-Mills-Chern-Simons theory. The basic argument involves a number of different steps.

\noindent\underline{Step 1}:

Consider the calculation of the expectation values of Wilson lines (which are the observables) in a Chern-Simons theory of level number $k$. We shall look at this calculation both in a Hamiltonian formulation and from the point of view of a functional integral approach. Consider first the Hamiltonian formulation on $\Sigma \times {\mathbb R}$ where, for the ensuing discussion, it is adequate to take the spatial manifold $\Sigma$ as ${\mathbb R}^2$ or the Riemann sphere
${\mathbb R}^2 \cup \infty$. In the $A_0 =0$ gauge, the wave functions must obey the Gauss law
condition
\beq
\delta_\epsilon \, \Psi (A ) = \int \epsilon^a~ \left[ {k \over 2\pi} \bdel A + \sum_r
(-it^a)_{(r)} \delta^{(2)} (x -x_r) \right]\, \Psi
\label{alt1}
\eeq
where $\delta_\epsilon \, \Psi$ denotes the change in $\Psi$ under the infinitesimal gauge transformation
$A \rightarrow A - D \epsilon$ and $t^a_{(r)}$ denotes the charge matrix for the $r$-th Wilson line.
We are using the holomorphic polarization for the wave functions
{\footnote{There is a slight change of notation compared to \cite{Bos-Nair}.
The present $z$, $\bz$ are the $\bz$ and $z$ of that paper.}}.
In the absence of any Wilson lines, this equation tells us that the ground state wave function
is given by
\beq
\Psi_0 =  \chi_0\, \exp ( k S_{wzw} (M))
\label{alt2}
\eeq
The constant $\chi_0$ is determined by the  normalization integral
\beq
\vert \chi_0\vert^2 \, \int d \mu (H) \, \exp \left[ (k + 2\, c_A) S_{wzw}(H)\right]
\,\, = \, 1
\label{alt3}
\eeq
In the exponent of the integrand, the term $k \, S_{wzw}(H)$ arises from the wave function,
and the remainder $2 c_A S_{wzw} (H)$ is the Jacobian for the conversion from the $A$'s to 
$H$.

\noindent\underline{Step 2}:

If we consider two charges, conjugate to each other, at positions $\vx_1$ and $\vx_2$, the solution to the Gauss law
(\ref{alt1}) is 
\beq
\Psi = \chi (z_1 , z_2 )\, M(1) \, M^{-1}(2) \, \Psi_0
\label{alt4}
\eeq
where $\chi$, which can depend on the coordinates of the charges, is to be determined by
the requirement that $\Psi$ should obey the Schr\"odinger equation. The Hamiltonian has the form
\beq
{\cal H} =  -i \sum_r \left[ {\dot\bz}_r {\delta \over \delta A(x_r)} + {\dot z}_r A(x_r) \right]
\label{alt5}
\eeq
The action of ${\cal H}$ on $\Psi$ produces singular terms via terms like
$\delta M(1) /\delta A(x_1)$. A regularization of this singularity shows two results:
\begin{enumerate}
\item There is a shift of $k \rightarrow \kappa = k + c_A$ in the expression (\ref{alt5}) for
${\cal H}$.
\item The Schr\'odinger equation becomes the Knizhnik-Zamolodchikov equation for the
chiral blocks with parameter $\kappa$. In other words, $\chi$ is
a chiral block for the level $k$ $SU(N)$ WZW theory.
\end{enumerate}

\noindent\underline{Step 3}:

We also know that $\Psi$ in (\ref{alt4}) should have a normalization integral independent of
$z_1$, $z_2$. The relevant integral is of the form
\beq
{\cal I} = \vert \chi\vert^2 \, \int d \mu (H)\, \exp[{\bar k} \,S_{wzw}(H)]~ 
H(1) \, H(2)^{-1}
\label{alt6}
\eeq
where we take the integration measure to be of the form $d\mu (H) \, \exp[{\bar k} S_{wzw}(H)]$.
We will leave ${\bar k}$ for this argument, even though explicit computation will show that
it is $k + 2 \, c_A$, as in (\ref{alt3}). Other than $\vert \chi\vert^2$, ${\cal I}$ is the correlator
$\la H(1) \, H(2)^{-1}\ra$ of a hermitian WZW theory (an $SL(N, {\mathbb C})/SU(N)$
theory) of level number ${\bar k}$. This correlator must exactly cancel the $z_i$-dependence of
$\vert \chi\vert^2$ to make the normalization of the state (\ref{alt4}) equal to a constant.
This, in turn, implies that $\la H(1) \, H(2)^{-1}\ra$ should obey the KZ equation
as in the $SU(N)$ WZW theory, but with $\kappa \rightarrow -\kappa$.
On the other hand, we also know directly from the path-integral for the hermitian WZW model, 
that the correlators of the level ${\bar k}$ hermitian model are the same as those of the
$SU(N)$ WZW model of level $-{\bar k}$. The consistency of these statements then requires that
\beq
- {\bar k} ~+~ c_A = - \left( k+ c_A \right)
\label{alt7}
\eeq
identifying ${\bar k} = k + 2 \, c_A$. Thus indirectly we identify the measure of integration
as $d\mu (H)\, \exp \left[ (k + 2\, c_A) S_{wzw}(H)\right]$. At this stage, we may take
$k \rightarrow 0$ to get the measure for gauge fields (with no contribution from the CS action).

\noindent\underline{Step 4}:

The shift $k \rightarrow \kappa = k + c_A$ in the Hamiltonian
(and hence in the KZ equation which is the Schr\"odinger equation for the CS theory)
can also be seen, in fact, more easily seen, in the path-integral approach.
We can, in principle, calculate observables (Wilson lines) by using a quantum effective action
$\Gamma$. Thus starting from the level $k$ Chern-Simons action, we first determine 
the corresponding $\Gamma$. It is well know that $\Gamma$ has the same form as the Chern-Simons action, but with $ k \rightarrow k + c_A$. All observables, from this method of calculation,
involves a single parameter $\kappa = k + c_A$. Since the same observables can also be calculated via the Hamiltonian approach, this tells us that the regularization of the Hamiltonian must produce the same shift $k \rightarrow k + c_A$. Taking this ingredient from the covariant path-integral approach,
we can bypass the regularization
procedure and identify the needed shift of $k$ in the Hamiltonian, and then combining this with the requirement of ${\cal I}$ being independent of the $z_i$'s, we have a method of identifying the measure of integration. 

\noindent The line of reasoning outlined above may be summarized as follows:
\begin{enumerate}
\item Identify the shift of $k$ in the effective action of the CS theory via a covariant
path-integral or Feynman diagram calculation.
\item The shifted $k$ from the effective action determines the shifted parameter
$\kappa$ in the Hamiltonian and hence in the KZ equation for the $SU(N)$  WZW theory.
\item The KZ equation (the Schr\"odinger equation) determines the $z_i$ dependence of
$\chi$ in the wave functions.
\item For the normalization to be independent of the $z_i$'s, the $H$-correlators
(calculated in the $SL(N,{\mathbb C})/ SU(N)$ WZW theory of level number ${\bar k}$)
 must obey the KZ equation of the $SU(N)$ WZW theory with parameter $- \kappa$.
 \item The chiral blocks of the $SL(N, {\mathbb C})/SU(N)$ WZW theory of level number
 ${\bar k}$ are the same as those for the $SU(N)$ theory with level $- {\bar k}$
 \item Putting these together, we find
 \beq
\left. \begin{matrix}
{\rm KZ~ parameter ~of}~ SU(N)~{\rm WZW}\\
{\rm ~~~~~~~~~~~~~~theory ~with ~level}~ - {\bar k}\\
 \end{matrix}\right\}
=
- \left\{ \begin{matrix}
{\rm KZ ~parameter ~of }~SU(N)~{\rm  WZW}\\
 {\rm theory~ with~ level}~ k~~~~~~~~~~~~~~~~\\
 \end{matrix}
 \right.
 \label{alt8}
 \eeq
\end{enumerate}
This last equation thus identifies the measure of integration. We can take $k \rightarrow 0$ at this stage to obtain the result for the gauge fields alone.

One may now wonder where the Yang-Mills action enters this line of reasoning.
Of course, we are concerned with the measure for the gauge fields, so the action, {\it per se}, does not play an important role. However, the covariant calculations in the CS theory which initiate this sequence of steps generally require a regulator. The Yang-Mills action may be taken as
a higher derivative regulator for the theory, so that all calculations may be thought of as being
carried out in the YMCS theory. The direct calculation of the integration measure
in the YMCS also yields ${\bar k} = k + 2 \, c_A$ \cite{KKN-CS}. 
The pure CS limit is obtained by taking the limit of large $e^2$; the shift of
$k$ is known to be independent of $e^2$. The pure Yang-Mills limit is obtained by taking
$k \rightarrow 0$.

Once the measure is known, the analogue of the mass parameter can be identified via the reasoning given in subsection \ref{volume-measure}. For the YMCS theory, this would give
$( k + 2 \, c_A) (e^2/ 4\pi )$. This value agrees also with the direct calculation obtained by expressing the Hamiltonian of the YMCS theory in terms of the gauge-invariant variables
$H$ \cite{KKN-CS}.

\subsection{Expectations for supersymmetric theories}

The line of reasoning in the last subsection can now be applied to the supersymmetric theory.
The calculation of the shift of the level in the ${\cal N}$-extended supersymmetric Yang-Mills-Chern-Simons theory has
 been done in \cite{lee-scs}.
(If the addition of the Yang-Mills term is viewed purely as a higher derivative regulator term,
notice that we do need a supersymmetric Yang-Mills action to obtain a regulator which preserves supersymmetry.)
The result is
 \beq
 k \rightarrow \left\{ \begin{matrix} 
 k+ c_A~~\hskip .2in& {\cal N} = 0\\
 k+ {c_A/2} \hskip .2in& {\cal N} = 1\\
 k~~~~~~~~~\hskip .2in& {\cal N} \geq 2\\
 \end{matrix}\right.
 \label{susy1}
 \eeq
Following equation (\ref{alt8}), we then find
that the inner product of wave functions for the supersymmetric YMCS theory should be
\beq
\la 1\vert 2\ra = \int d\mu(H)\, \exp[ {\bar k} \, S_{wzw}(H)]~
d\,[{\rm Fermions}]~ \Psi_1^* \, \Psi_2
\label{susy2}
\eeq
where
\beq
{\bar k} =  \left\{ \begin{matrix} 
 k+ 2\, c_A~~\hskip .2in& {\cal N} = 0\\
 k+ {c_A}~~~ \hskip .2in& {\cal N} = 1\\
 k~~~~~~~~~\hskip .2in& {\cal N} \geq 2\\
 \end{matrix}\right.
 \label{susy3}
 \eeq
with the corresponding mass gap ${\bar k} \,(e^2/ 4\pi)$. For the nonsupersymmetric YMCS theory, i.e., for ${\cal N} = 0$, this means  we expect a mass gap as we have seen before, which survives to the pure YM limit, when $k$ is taken to zero. For ${\cal N} = 1$, we cannot take $k =0$ because of the parity anomaly; $k =1 $ is the smallest value possible.
(In addition to general anomaly-based arguments, there is also evidence from lattice simulations for the need for the Chern-Simons term \cite{n=1-lattice}.)
In this case, we will have a nonzero value for ${\bar k}$ and hence a mass gap for the theory, although of a magnitude different from the pure YM case.
For ${\cal N} \geq 2$, the result (\ref{susy3}) is consistent with the absence of a mass gap.
 
 As mentioned in the introduction, these expectations also agree with other analyses in the literature. 
 First of all, for ${\cal N} = 8$, there is the relation to the ABJM theories; namely that the maximally supersymmetric theory is expected to flow to a conformal theory\cite{abjm} in the infra-red.  A necessary condition for the flow to conformality  is the massless nature of the physical spectrum of Yang-Mills theory. 
 Again, for the ${\cal N} =4 $ theories, 
 the constraints of unbroken  supersymmetry prevent a mass term \cite{seiberg-witten}, while 
partial gauge symmetry breaking can occur giving rise to a Coulomb branch.
Also for ${\cal N} = 2$ theories, the expectation is that there is no mass gap,
but with no stable supersymmetric vacuum \cite{seiberg-witten, N=2-vac}.
The absence of mass gap 
 for ${\cal N} = 2$ has also been analyzed by different methods in \cite{N=2-brane, unsal-bion}. 
 
The issue of the ground state, or the lack thereof, for the ${\cal N} = 2$ theory is a very interesting one,
but is beyond what the measure by itself can address. Nevertheless, recall that $S_{wzw}(H)$ in the exponent of the measure is crucial for
 the convergence of volume integrations. This is true if we start with
 $\Psi_0 =1 $ as the eigenstate of the kinetic energy operator and improve on it, as done in
 \cite{KKN-long}, or in evaluating expectation values using the chiral boson version as done in
 \cite{KKN-string}. The absence of such a factor for ${\cal N} \geq 2$ also points in the direction of
 an unstable vacuum.
 
We have imposed supersymmetry in obtaining these answers. Supersymmetry breaking 
for ${\cal N} = 1$ is another important question, but, again, 
 the measure calculation by itself is not adequate for analyzing this. However, further development along the lines outlined here, with a calculation of the ground state wave function, could possibly shed light on this matter. 

We will now turn to the more explicit calculation of the measure of integration and the Hamiltonian in the supersymmetric theories.
\section{Measure from a chiral anomaly computation}\label{fermi-measure}
We start by considering Majorana fermions in the adjoint representation of the group $SU(N)$
in $2+1$ dimensions.
The conventions for the gamma matrices is
\beq
\gamma ^\mu = \{ i\sigma ^3, \sigma ^1, \sigma ^2\}: \hspace{.3cm} C\gamma ^\mu C^{-1} = -(\gamma ^\mu)^{T}
\label{anom1}
\eeq
where $C$ is
the charge conjugation matrix which may be taken as $C = \gamma ^2  = \sigma ^2$. The Majorana fermions $\Psi$ satisfying $\bar \Psi = \Psi^T\, C$ can be brought to the form
\beq
 \Psi = \begin{pmatrix}
\psi\\ \psi^\dagger \end{pmatrix}
\label{anom2}
\eeq
signaling that there is a single field degree of freedom.

Because of the Majorana condition, the gauge-invariant parametrizations for 
 $\psi $ and $\psi^\dagger$ must be related.  There are two choices possible for the gauge-invariant fermionic variables $\chi$ and $\chi^\dagger $ compatible with the Majorana condition on $\Psi$. These are given by
\begin{align}
{\rm Choice~I}:\hskip .3in&
\begin{pmatrix}
\chi^a\\ \chi^{a \dagger} \end{pmatrix} = \begin{pmatrix}
(M^{-1})^{ab} \psi^b\\ (M^\dagger )^{ab}\psi^{b \dagger} \end{pmatrix}\nonumber\\
{\rm Choice~ II}:\hskip .3in&  \begin{pmatrix}\chi^a\\ \chi^{a \dagger} \end{pmatrix} = \begin{pmatrix}
(M^{\dagger })^{ab} \psi^b\\ (M^{-1})^{ab}\psi^{b \dagger} \end{pmatrix}
\label{chi}
\end{align}
Here $M^{ab} = 2\, \Tr(t^a \, M\, t^b\,M^{-1})$ is the adjoint representative of $M$.
Denoting $M = e^{t^a\theta ^a}$, for small $\theta^a$, these two choices 
become
\beq
\begin{pmatrix}
\chi^a\\ \chi^{a \dagger} \end{pmatrix}  = \left[ 1 \mp i\, {\rm Im}(\theta)^aT^a \mp {\rm Re}(\theta)^aT^a\gamma^5  + \cdots \right]\, \Psi \label{anom3}
\eeq
In this expression, $\gamma ^5$ is nothing but $\sigma ^3$. We call it $\gamma ^5$
because it is the chirality matrix in the two-dimensional sense, for the spatial gamma matrices used above. The dots represent higher order terms in $\theta$. The upper and lower signs correspond to the choices I and II respectively.  
For the transformation of the measure of integration for the fermion fields, we find,
\beq
d\psi\,d\psi^\dagger  = \mbox{Det}\left[ 1 \mp i \,{\rm Im}(\theta)^aT^a \mp {\rm Re}(\theta)^aT^a\gamma^5 + \cdots \right] ~d\chi \,d\chi^\dagger
\label{anom4}
\eeq 
The evaluation of the determinant will require regularization. One possibility if to use
the well known techniques of chiral anomaly computations in (Euclidean) two dimensional spacetime.  For example, we can regulate the determinant using $e^{-(\gamma\cdot D) ^2/M^2}$, where $M$ is the cut-off and $\gamma\cdot D$ is the hermitian Dirac operator in two dimensions
given by
\beq
\gamma \cdot D = \sum_{i=1,2} i\gamma ^i\partial_i + \gamma^i\, T^aA^a_i
\label{anom5}
\eeq
The computation of the regularized determinant  is then standard and gives,
upon taking $M\rightarrow \infty $,  
\beq
\mbox{Det}\left[ 1 \mp i\, {\rm Im}(\theta)^aT^a \mp {\rm Re}(\theta)^aT^a\gamma^5 + \cdots \right] = \mp \frac{c_A}{2\pi}\int F^a_{12}\,  {\rm Re}(\theta)^a 
\label{anom6}
\eeq
The above equation is valid to linear order in  $\theta$. 
We can find the determinant for finite $\theta^a$ by first noting the the variation of the WZW action
(\ref{vol3}) gives
\beq
\delta S_{wzw}(H) = -\frac{1}{2\pi} \int  F^a_{12}\, {\rm Re}(\theta)^a\label{variationwzw}
\eeq
Comparing this with (\ref{anom6}), we see that we can integrate the latter to obtain
\beq
\mbox{Det}\left[ 1 \mp i \,{\rm Im}(\theta)^aT^a \mp {\rm Re}(\theta)^aT^a\gamma^5 + \cdots \right] = 
\exp \left[ \pm c_A\, S_{wzw}(H)\, \right]
\label{anom7}
\eeq
 
A few remarks are in order at this stage. The first is concerned with the regularization.
It is not {\it a priori} clear that we should regulate the determinant in the same way as is done for the chiral anomaly in two dimensions.
If we have an intrinsically two-dimensional theory, then there is an independent way to
corroborate this result.
We can consider the change in the action, which is related to the conservation of the axial current and, from working out this conservation law using Feynman diagrams, it is easy enough to reproduce the same result as obtained by the regularization of the Jacobian as outlined above.
But for the $(2+1)$-dimensional case, the argument is more involved. The fermionic action is of the first-order in the time derivatives. Its quantization can be done in terms of fermionic coherent states. The normalization of the states then involves the K\"ahler potential associated to the symplectic one-form.
In the present case, it is of the form $\psi^\dagger \psi$. In constructing the functional integral via dividing up the interval $\tau$ in $\la 0\vert e^{-\tau {\cal H}} \, \vert 0\ra$, the inner product
$\la \psi_n \vert \psi_{n+1}\ra \sim \exp ( \psi^\dagger_n \psi_{n+1} ) \approx
\exp (  \psi^\dagger_n \psi_n +  \psi^\dagger_n {\dot \psi}_n \epsilon )$
combines with the term involving the K\"ahler potential in the measure
$\exp (- \psi^\dagger_n \psi_n)$ and $\exp (- \epsilon {\cal H})$ to form the three-dimensional action.
(Similar arguments hold for the bosonic coherent states as well.) In other words, the factor
involving the K\"ahler potential in the measure is exactly what we would expect from 
writing out the action as a sum over discrete time-intervals $\epsilon$.
Now, if the three-dimensional theory generates a Chern-Simons term,
then the corresponding K\"ahler potential being $S_{wzw}(H)$, we would expect such a term to be generated in the measure of integration for the inner product of states.
We know from \cite{pisarski-rao, lee-scs} that there is a level shift of the CS term induced in
three dimensions.
So if we choose a regularization that agrees with the
three-dimensional calculation, and hence with supersymmetry,
then we should get the result (\ref{anom7}). The result is dependent on whether supersymmetry is preserved, just as it is in the three-dimensional covariant calculation.

The second observation is about which of the choices in (\ref{chi}) is the right one.
Again, it is a matter of which symmetry is to be preserved. We shall see that supersymmetry requires the second choice of the variables.

These considerations suggest a different strategy  for the quantization of the theory. We can set up the supercharges and then obtain the Hamiltonian from the anticommutator of the supercharges,
guaranteeing a supersymmetry-preserving regularization of the terms in the Hamiltonian.
We now turn to this task.

\section{Mass-gap and Measure for $\mathcal{N} = 1$ YMCS Theory}

In this section we shall focus on the Hamiltonian formulation of the minimally supersymmetric gauge theory in $D=2+1$. The measure of integration for the inner product
will emerge as one of the results of this analysis.
Our analysis will be general enough to incorporate Chern-Simons couplings  and the results pertaining to the integration measure
will turn out to be consistent with the discussion in the previous section. For the sake of completeness, we start with a brief overview of the canonical quantization of the $\mathcal{N} = 1$ Yang-Mills-Chern-Simons system.

\subsection{Canonical Quantization of the $\mathcal{N} = 1$ Theory}\label{canonical}

When the Chern-Simons level number is set to zero, the action for the $\mathcal{N} =1 $ supersymmetric Yang-Mills theory is given by
\beq
S = -\frac{1}{4\,e^2}\int F^a_{\mu \nu}F^{a\mu \nu} - \frac{i}{2e^2} \int \bar{\Psi}^a(\gamma ^\mu D_\mu \Psi)^a\label{SYM}
\eeq
This action is invariant under the supersymmetry transformation
\beq
\delta_\epsilon A_\mu^a = -i \, {\bar\epsilon}\, \gamma_\mu \Psi^a, \hspace{.3cm} \delta_\epsilon \Psi^a =  \frac{1}{2}F^a_{\mu \nu }\gamma^{\mu \nu}\epsilon\label{n1susy}
\eeq
In a Hamiltonian setting, the supercharges which generate this transformation are
\beq
Q^\dagger  =\int( i\Psi^\dagger \gamma^i \frac{\delta}{\delta A^i} + \frac{1}{e^2}\psi^\dagger B), \hspace{.3cm} Q  =\int( i \gamma^i \Psi \frac{\delta}{\delta A^i} + \frac{1}{e^2}\psi B)
\label{canon3}
\eeq
As with any other fermionic observable in the theory, the supercharge, is a two-component spinor,
$Q^1 = q$, $Q^2 = q^\dagger$.
Using the canonical anticommutation relation
\beq
\{\Psi^\dagger _\alpha, \Psi_\beta\} = e^2\delta _{\alpha \beta}\label{canon4}
\eeq
it is readily verified that $\{Q^\dagger_\alpha, Q_\beta\} = 2 \, {\cal H}\,\delta_{\alpha \beta}$, where the $A_0 = 0$ gauge choice is implied.

The classically massless theory described leads to an anomalous, and hence inconsistent, quantum theory. The parity anomaly of the fermionic action forces the partition function of this theory to be trivial \cite{n=1-lattice, seiberg-witten}.
To obtain a consistent theory with minimal supersymmetry in three spacetime dimensions,
one must necessarily add a supersymmetric Chern-Simons term to the action,
\beq
S_{SCS} = -\frac{k}{4\pi} \int d^3x ~\Tr \left[\left(A_\mu\partial_\nu A_\alpha - \frac{2}{3}A_\mu A_\nu A_\alpha\right)\epsilon^{\mu \nu \alpha} + ie^2\bar{\Psi}\Psi\right]\label{SCS}
\eeq
 $S+S_{SCS}$ is also invariant under (\ref{n1susy}). In what is to follow, we shall consider the theory with the Chern-Simons term added for the sake of generality \cite{DJT}. The electric field operators, and the commutation relations between them  are given by
\beq
\begin{split}
E = -i \frac{\delta}{\delta \bar A} + &\frac{ik}{4\pi} A, \hspace{.3cm} \bar{E} = -i \frac{\delta}{\delta  A} - \frac{ik}{4\pi} \bar{A}\\
&[E,\bar E] = -\frac{k}{2\pi}
\end{split}\label{canon5}
\eeq
The Gauss law for the theory, which is a constraint on the physical states, is
\beq
(D\bar \Pi)^a + (\bar D \Pi)^a  + \frac{ik}{4\pi} (\partial \bar A^a - \bar \partial A^a) - \frac{i}{2} f^{abc} \Psi ^{\dagger b}\Psi ^c = 0
\label{canon6}
\eeq
Wave functions, $\Xi$, satisfying the Gauss law are of the form
\beq
\Xi  = \exp \left( \frac{k}{2} \bigl[ S_{wzw}(M^\dagger) - S_{wzw}(M)\bigr] \right)\Lambda '(H, \chi, \chi^\dagger) = e^{i\omega}\Lambda
\label{canon7}
\eeq
Since $e^{i\omega }$ is a pure phase, it does not change the norm of the wave functionals, however it can affect the matrix elements of dynamical quantities. This can be seen explicitly by writing the observables as operators on $\Lambda$ rather than on $\Xi$.
For example, the effective supercharge acting on  $\Lambda$  is given by 
$\widetilde q = e^{-i\omega}qe^{i\omega}$.
The precise form of the transformed supercharge is
\beq
\widetilde q = \int\left[i (\psi^a)^\dagger  \left(\frac{\delta}{\delta A^a} +\frac{k}{4\pi}(\bar A - \bar a)^a\right)  + \frac{1}{e^2}\psi^a  B^a\right]\label{charge}
\eeq
where $a$ and $\bar a$ are auxiliary fields satisfying the `eikonal' equations
\beq
\bar D a = \partial \bar A, \hspace{.5cm} D\bar a = \bar \partial A 
\label{canon8}
\eeq
We may solve for $a$, ${\bar a}$ explicitly in terms of the matrix parametrization (\ref{vol1}) as
\beq
\bar a = -\bar \partial M M^{-1}, \hspace{.5cm} a = M^{\dagger -1} \partial M^\dagger
\label{canon9}
\eeq
It is instructive to note the form of the Hamiltonian as well. A direct evaluation of the 
anticommutator of supercharges  gives
\beq
\begin{split}
\widetilde{\mathcal{H}} = \frac{1}{2} \,{\{\widetilde q, \widetilde q ^\dagger\}}= &-\frac{e^2}{2}\int \frac{\delta ^2}{\delta A^a \delta \bar A ^a} + \frac{m_k}{2}\int \left[ (A-a)^a\frac{\delta }{\delta A^a} - (\bar A - \bar a)^a\frac{\delta}{\delta \bar A^a}\right]\\
&+ \frac{1}{2e^2}\int \left[ m_k^2 (A-a)^a  (\bar A - \bar a)^a +  B^aB^a\right]
+  i \int \bar \Psi (\gamma ^i  D_i- m_k)\Psi
\end{split}\label{H}
\eeq
$m_k =(e^2 k /4\pi )$ is the topological mass. 
(There can also be a dynamical generated addition to this mass which requires a more careful analysis.)
Further, we have supersymmetry, i.e.,
$[\widetilde q, \widetilde{\mathcal{H}}] = 0$.
This is easily checked using the 
commutation rules
\beq
\begin{split}
&[\frac{\delta}{\delta A^a(x)}, \bar a^b(y)] = F^{ab}(x,y) \equiv -M^{ac}(x)\frac{1}{\pi(\bar x - \bar y)^2}M^{bc}(y),\\
&[\frac{\delta}{\delta A^a(x)},  a^b(y)] = 0, 
\end{split}
\eeq
and the
symmetry relation $F^{ab}(x,y) = F^{ba}(y,x)$.

\subsection{Choice of Gauge Invariant Variables and $\mathcal{N} = 1$ Supersymmetry}

We now turn to  the choice of gauge invariant variables. As noted previously, there are two different ways of constructing gauge-invariant variables from the adjoint Majorana fields.
However a unique choice is picked out by analyzing the supercharge. To see this, we first recall that the change of variables from $A \rightarrow M$ implies that operator corresponding to the antiholomorphic component of the electric field and the magnetic field are given by \cite{KKN-long}
\beq
-i\frac{\delta}{\delta A^a} = -iM^{ab}(x)\int_y \mathcal{G}(x,y)p^b(y), \hspace{.3cm} B^a = -\frac{2\pi}{c_A}(M^{\dagger -1})^{ab}\bar \partial J^b
\eeq
where the gauge invariant generator $p^b$ acts as a $SU(N)$ rotation generator on $M$ (see equation \ref{pcomm}) and $\mathcal{G}$ is a regularized expression for the Greens' function for the holomorphic derivative $\partial$; the explicit expression for $\mathcal{G}$ can be found in the appendix. Using these relations we see that
 the formula for the supercharge (\ref{charge}) becomes
 \beq
 \begin{split}
\widetilde q &= \int\left[i (\psi^a)^\dagger  \left(\frac{\delta}{\delta A^a} +\frac{k}{4\pi}(\bar A - \bar a)^a\right)  + \frac{1}{e^2}\psi^a  B^a\right]\\
&=  i\int_x \underbrace{\psi^{a^\dagger}(x) M^{ab}(x)}_{\chi^{\dagger b}}\left[\int_y \mathcal{G}(x,y)p^b(y)  + \frac{k}{4c_A}\bar{J}^b(x)\right] - \frac{1}{e^2}\frac{2\pi}{c_A}\int \underbrace{\psi^a (M^{\dagger -1})^{ab}}_{\chi^b}\, \bar \partial J^b\label{choicen1}
\end{split}
\eeq
  In the above equation we have used the following definitions for the currents:
\beq
J = \frac{c_A}{\pi}\partial H H^{-1} \hspace{.3cm} \bar J = \frac{c_A}{\pi} H^{-1}\bar \partial H
\eeq
The second line in (\ref{choicen1}) tells us that the natural fermionic variable is $\chi^\dagger(x) = M^{-1}\psi^\dagger $ (and its hermitian conjugate), which was precisely our choice II which $reduced$ the volume of configuration space by a factor of $e^{-c_AS_{wzw}(H)}$. From the point of view of the renormalization of the level number, the fermionic contribution removes half of the contribution to the renormalized level number that one had in the pure glue theory, which is consistent with previous perturbative and nonperturbative computations,
as we have already discussed.
Thus if the only change of variables we make is $A \rightarrow M$ and let the choice of the gauge invariant fermionic variable be dictated by the form of the supercharge, then choice II is the unique answer for $\mathcal{N}=1$ SUSY. 
We will see later that similar reasoning can be used to show
that
the fermionic contribution to the measure exactly cancels the measure factor obtained the case of the pure glue theory, for $\mathcal{N}=2$ and $\mathcal{N}=4$ supersymmetry.

 Starting from (\ref{choicen1}), our strategy for deriving the Hamiltonian is to compute the adjoint of the supercharge and then anticommutator  between the two.
As in the case of the $\mathcal{N} = 0$ theory \cite{KKN-CS},  it is convenient to  define the wave functions $\Phi$ such that 
\beq
\Lambda^\prime = \exp\left( {\frac{k}{2}S_{wzw}(H)}\right) ~\Phi
\label{Lambda}
\eeq
The integration measure for the inner product for the $\Phi$'s now has the form
\beq
d\mu = d\mu (H)\,  \exp \left[ (k + (2-n)c_A) \, S_{wzw}(H)\right]
\label{measure}
 \eeq
 The factor of $2\,c_A$ in the exponent is the contribution relevant to the pure glue theory. 
 Based on the arguments given in section 2,  we expect $n = 0$ for $\mathcal{N} = 0$ and   $n = 1$ for $\mathcal{N} = 1$.  However, we shall keep $n$ as arbitrary for now. After 
 absorbing the factor $\exp\left( \frac{k}{2}S_{wzw}(H) \right)$ into the wave functions as in
 (\ref{Lambda}), we find, for the supercharge, as an operator on the $\Phi$'s, 
 \beq
 q' = i\int \chi^{\dagger a}(\mathcal{G}p)^a - \frac{1}{e^2}\frac{2\pi}{c_A}\int \chi ^a \bar\partial J^a\label{qkkn}
 \eeq
 The adjoint supercharge has to be computed with respect to the measure (\ref{measure}), and
 is found to be
 \beq
 q^{\prime \dagger} =  - i \int \chi ^a\left((\bar{\mathcal{G}}\bar{p})^a - i\frac{k}{2\pi}(\partial HH^{-1})^a + i \frac{nc_A}{2\pi}(\partial HH^{-1})^a\right) - \frac{1}{e^2}\frac{2\pi}{c_A}\int (\chi ^a \bar\partial J^a)^\dagger\label{adqkkn}
 \eeq 
 $\bar p$ acts as the $SU(N)$ generator on $M^\dagger$ via left action \cite{KKN-long}.
 The term proportional to $k$ on the right hand side comes from $p$ acting on $ e^{kS_{wzw}(H)}$ in the computation of the adjoint. The term proportional to $c_A$ is what remains of  the action of $p$ on $\chi ^\dagger $ (which induces a contribution from $\mathcal{G}(0)$) and the action on  $e^{(2-n)c_AS_{wzw}(H)}$.)\footnote{For details about the regularization scheme leading to a finite answer for the coincident limit of $\mathcal{G}$, see \cite{KKN-long}, and for a discussion about the compatibility of the regulated expression with the measure on $\mathcal{C}$, see \cite{robustness}.}
 A naive evaluation of  the anticommutator of the supercharges gives the Hamiltonian as
\begin{align}
{\mathcal H} = \frac{1}{2} \{q^\prime, q^{\prime \dagger}\} = & \frac{e^2}{2} \int_{u,v} \Pi^{rs}(u,v) \bar p^r(u) p^s(v) + \frac{2\pi^2}{e^2\, c_A^2}\int(\bar \partial J^a\, \bar \partial J^a)\nonumber\\
&+ \frac{e^2(k-nc_A)}{4\pi}\int J^a \frac{\delta}{\delta J^a}\label{naiveham}\\
&-\frac{1}{e^2}\int (\chi ^\dagger \bar{\mathcal{D}}_{\bar{J}}\chi ^\dagger - \chi \mathcal{D}_J\chi) + \frac{1}{e^2}\left(\frac{ke^2}{4\pi} - \frac{nc_Ae^2}{4\pi}\right)\int \chi^{\dagger a} (H^{-1})^{ab}\chi^b
\nonumber
\end{align}
 The formula above can be understood as follows.
 The first line is the  pure Yang-Mills Hamiltonian, where on states formed out of the bosonic $J$ fields alone, the kinetic energy operator is
 \begin{align}
 T_{YM} = &\frac{e^2}{2} \int_{u,v} \Pi^{rs}(u,v) \bar p^r(u) p^s(v) \equiv \frac{e^2c_A}{2\pi}\left(\int J^a\frac{\delta}{\delta J^a} + \int \Omega ^{ab}(xy)\frac{\delta}{\delta J^a(x)}\frac{\delta}{\delta J^b(y)}\right)\nonumber\\
& \hskip .4in\Omega ^{ab}(x,y) = \frac{c_A}{\pi^2}\frac{\delta^{ab}}{(x-y)^2} - i f^{abc}\frac{J^c(y)}{\pi(x-y)}
\end{align}
Notice that $\Omega^{ab}(x,y)$  is the two-point function appearing in the operator product expansion of $J^a(x)\, J^b(y)$ in the WZW model. The second line of (\ref{naiveham})  is the contribution to the mass-gap from the factors of $\exp \left[ (k -n \,c_A) \, S_{wzw}(H)\right]$ in the measure. This contribution vanishes when $k=n=0$ in the pure Yang-Mills case, as it should. The last line in the Hamiltonian is the fermionic contribution where $\mathcal{D}$ and $\bar{\mathcal{D}}$ are the holomorphic and anti-holomorphic covariant derivatives respectively with $J$ 
and $\bar{J}$ playing the roles of the connections.
In deriving these results we have also used the fact that, on functionals of $J$,
 \beq
 (\mathcal{G}p)^a = i\, \frac{c_A}{\pi}\, H^{-1}\, \frac{\delta}{\delta J}
 \eeq
 
The expression (\ref{naiveham}) does not seem to be the correct expression for the Hamiltonian as the bosonic and fermionic masses in $\mathcal{H}$ do not appear to be equal. Looking at the terms that will become mass terms when the Hamiltonian is truncated to the quadratic level, the mass for the bosons is given by the term
 \beq
{\mathcal H}_{bos-mass} =  \frac{e^2}{4\pi}(k + [2-n]c_A)\int J^a\frac{\delta}{\delta J^a}
\label{bos-mass}
 \eeq
corresponding to the value $m_b =  (k + [2-n]c_A) (e^2/4\pi)$.
On the other hand the fermionic mass is
$ m_f = (k  -n)c_A \, (e^2/4\pi)$.
Without the equality of the masses the Hamiltonian is obviously not supersymmetric. The resolution actually lies in the use of the Gauss law. Let us eliminate $E$ from (\ref{adqkkn}) using the Gauss law, which in the original variables was given by
 \beq
 \mathcal{I}^a = \left(\bar D \frac{\delta }{\delta \bar A} + D \frac{\delta }{\delta A}\right)^a + \frac{1}{e^2}f^{abc} \psi^{b\dagger}\psi^c = 0 ,
 \label{gausslaw1}
 \eeq
 or equivalently, on physical states,
 \beq
 \bar p^a = (Hp)^a + \frac{1}{e^2}f^{alm}(H\chi^\dagger)^l\chi^m
 \eeq
 The fact that the Gauss law gives a relation between ${\bar p}^a$ and
 $p^a$ when they act on gauge-invariant functionals was already noted in
 \cite{KKN-long}. Since we have functionals of $J$, rather than ${\bar J}$, it is preferable to
 eliminate ${\bar p}^a$.
 This elimination of ${\bar p}^a$ (or $E$) in the expression for the adjoint of the supercharge
 involves
 \beq
 -\psi^a \frac{\delta }{\delta \bar{A}^a} = \frac{1}{e^2}\int_y \psi^a (x) (\bar{D}^{-1})^{ab}(x,y) f^{bmn} \psi^{m \dagger}(y)\psi^n(y) + \cdots
 \eeq
 The ellipsis represent other terms that we shall come back to shortly. Now, the normal ordering of
 this expression results in the term
 \bsp
  -\psi^a \frac{\delta }{\delta \bar{A}^a} &=   (\bar{D}^{-1})^{mb}(x,x) f^{bmn} \psi^n(x) + \cdots\\
  &= -i(M^{\dagger -1})^{nl} J^l\psi^n + \cdots
 \end{split}
 \label{correction1}
 \eeq
 where we have used the regularized result \cite{KKN-long},
 \beq
(\bar{D}^{-1})^{ab}(x,x)  = \frac{i}{c_A}f^{abc}(M^{\dagger -1})^{cl} J^l
 \eeq
 Obviously, the term (\ref{correction1}) in $q'^\dagger$, when anti-commuted with the
 term $\int \chi^{a \dagger} (\mathcal{G}p)^a$  in $q'$ produces both a bosonic and a fermionic mass-term. This will  restore the equality of the masses as required by supersymmetry.
 The complete normal ordered expression that results from the elimination of $E$ in the adjoint supercharge is given by
 \bsp
 -\psi^a  \frac{\delta }{\delta \bar{A}^a}  =   \, &(\bar{D}^{-1})^{ab}(x,x) f^{ban}\psi^n(x) - \frac{1}{e^2}\int_y  (\bar{D}^{-1})^{ab}(x,y) f^{bmn}\psi^{m \dagger}(y) \psi^a(x) \psi^n(x) \\
& + \psi^a(x)\int_y  (\bar{D}^{-1})^{ab}(x,y) \left(D\frac{\delta}{\delta A}\right)^b
 \end{split}
 \eeq
 Or, in terms of the gauge invariant variables, this is equivalent to
 \bsp
 -i\int \chi^a(x)(\bar{\mathcal{G}}\bar p) ^a(x) = &-i\int \chi^a(x)(\bar{\mathcal{G}}H p) ^a(x) - \int \chi^a(x)J^a(x) \\
 &+ \frac{i}{e^2}\int \bar{\mathcal{G}}^{ab}(x,y) f^{blm}(H\chi^\dagger)^l(y)\chi^a(x)\chi^m(y)
 \end{split}
 \eeq
 Computing the Hamiltonian via the anticommutator of the supercharges we now get
  \bsp
\mathcal{H} = \frac{1}{2} \{q^\prime, q^{\prime \dagger}\} =& \frac{e^2c_A}{2\pi}\left(\int J^a\frac{\delta}{\delta J^a} + \int \Omega ^{ab}(xy)\frac{\delta}{\delta J^a(x)}\frac{\delta}{\delta J^b(y)}\right)
+ \frac{2\pi^2}{e^2c_A^2}\int(\bar \partial J^a \bar \partial J^a)\\
&+ \frac{e^2(k-nc_A)}{4\pi}\int J^a \frac{\delta}{\delta J^a} - \frac{ic_A}{2\pi}\int f^{ngz}\bar{G}(x,y)H^{zs}(y)\chi^{s \dagger}(y)\chi^g(y) \frac{\delta}{\delta J^n(x)}\\
&-\frac{1}{e^2}\int (\chi ^\dagger \bar{\mathcal{D}}_{\bar{J}}\chi ^\dagger - \chi \mathcal{D}_J\chi) + \frac{1}{e^2}\left(\frac{c_Ae^2}{2\pi} + \frac{ke^2}{4\pi} - \frac{nc_Ae^2}{4\pi}\right)\int \chi^{\dagger a} (H^{-1})^{ab}\chi^b\label{invh1}
\end{split}
 \eeq
 The equality between the bosonic and fermionic masses at the quadratic level is now manifest. The expression above is the correct gauge-invariant form of the Hamiltonian corresponding to the $\mathcal{N} = 1$ Yang-Mills-Chern-Simons theory. 
 
 \subsection{Checking SUSY Invariance of the $\mathcal{N}=1$ Hamiltonian} 
 
 Since the form of the  gauge invariant $\mathcal{N}=1$ Hamiltonian (\ref{invh1}) is rather different from the original Hmailtonian that one started with, containing several nonlocal terms, for instance,
as a further check on the calculations, we shall now verify its
supersymmetry invariance from first principles.
To see the invariance of (\ref{invh1}) under (\ref{qkkn}) we first note that the `mass-terms' are supersymmetric on their own.
\beq
[q', \int \left(J^a\frac{\delta}{\delta J^a} + \chi^{\dagger a} (H^{-1})^{ab}\chi^b\right)] = 0
\eeq
Before proceeding further, it is useful to rewrite various terms in the nonlocal parts of the Hamiltonian such that their commutators with $q'$ involve as few operator commutators as possible.  Denoting these two terms by $T_1$, $ T_2$, we write
\begin{align}
T_1 &\equiv   \frac{e^2c_A}{2\pi}\left( \int \Omega ^{ab}(xy)\frac{\delta}{\delta J^a(x)}\frac{\delta}{\delta J^b(y)}\right) \nonumber\\
& = \frac{e^2}{2}\left((\bar{D})^{-1ac}(xz)(DM\mathcal{G}p)^c(z)(M\mathcal{G}p)^a(x)\right)\label{T1T2}\\
T_2 & \equiv - \frac{ic_A}{2\pi}\int f^{ngz}\bar{G}(x,y)H^{zs}(y)\chi^{s \dagger}(y)\chi^g(y) \frac{\delta}{\delta J^n(x)} \nonumber\\
&=  \frac{1}{2}\int (\bar D)^{-1ac}(x,z) f^{cmn}(M\chi^\dagger)^m(z)(M^{\dagger -1}\chi)^n(c)(M\mathcal{G}p)^a(x)\nonumber
\end{align}
where the covariant derivative $D = \partial - \partial M M^{-1}$ and we have also used,
\beq
\bar{D}^{-1ab}(x,y) = M^{\dagger -1ac}(x)\, \bar{\mathcal{G}}(x,y)\, M^{\dagger cb}(y) \rightarrow M^{\dagger -1ac}(x)\frac{1}{\pi(x-y)}M^{\dagger cb}(y) 
\eeq
The expression on the r.h.s corresponds to the Green's function with the regulator removed.\\
Now we note that the sum $T_1+ T_2$ commutes with the part of the 
supercharge involving $p^a$; i.e., 
\beq
[i\int \chi^{\dagger a}(\mathcal{G}p)^a, T_2] = -\frac{ie^2}{2}\int  (\bar D)^{-1ac}(x,z)f^{cmn}(M\chi^\dagger)^m(z)(M\mathcal{G}p)^n(z)(M\mathcal{G}p)^a(x)
\eeq
The only commutator appearing above is between $\chi^\dagger $ and $\chi$. 
We also have
\beq
[i\int \chi^{\dagger a}(\mathcal{G}p)^a, T_1]  = +\frac{ie^2}{2}\int  (\bar D)^{-1ac}(x,z)f^{cmn}(M\chi^\dagger)^m(z)(M\mathcal{G}p)^n(z)(M\mathcal{G}p)^a(x)
\eeq
The only nontrivial commutator in this  calculation is between $\chi^{\dagger a}(\mathcal{G}p)^a$ and $-\partial MM^{-1}$ contained in $D$ in $T_1$. From these two results, we see that
\beq
[i\int \chi^{\dagger a}(\mathcal{G}p)^a, T_1 + T_2] = 0\label{cancel1}
\eeq
Moving on to the commutator of these terms with the rest of the supercharge, we obtain
\bsp
\Bigl[-\frac{2\pi}{e^2c_A}\int \chi^a\bar \partial J^a, T_2 \Bigr] = &+\frac{i}{e^2}\int f^{kmn}(M^{\dagger -1}\chi)^k(M\chi^\dagger)^m(M^{\dagger -1}\chi)^m\\
&-\frac{\pi}{c_A}\int  (\bar D)^{-1ac}(x,z)f^{cmn}(M^{\dagger -1}\bar \partial J)^m(z)(M^{\dagger -1}\chi)^n(z)(M\mathcal{G}p)^a(x)\label{nonlocal1}
\end{split}
\eeq
The first term on the right hand side involved the functional derivative with respect to $J$ in $T_2$ acting on the supercharge, while the second term is the result of the fermions anticommuting between the two terms.
Turning to the commutator with $T_1$,
\begin{align}
\Bigl[-\frac{2\pi}{e^2c_A}\int \chi^a\bar \partial J^a, T_1\Bigr]  &= i\int  (M^{\dagger -1}\chi)^m(D M\mathcal{G}p)^m \nonumber\\
&+ i \int (M^{\dagger -1}\chi)^l(x)(\bar{D}(x)D(x) \bar{D}^{-1}(x,y))^{la}(M\mathcal{G}p)^a(y)\nonumber\\
& =2\, i\, \int  (M^{\dagger -1}\chi)^m(DM\mathcal{G}p)^m \nonumber\\
&+ i \int (M^{\dagger -1}\chi)^l(x)([\bar{D},D](x) \bar{D}^{-1}(x,y))^{la}(M\mathcal{G}p)^a(y)\nonumber\\
&= 2\, i\,\int  (M^{\dagger -1}\chi)^m(DM\mathcal{G}p)^m \nonumber\\
&+\frac{\pi}{c_A}\int  (\bar D)^{-1ac}(x,z)f^{cmn}(M^{\dagger -1}\bar \partial J)^m(z)(M^{\dagger -1}\chi)^n(z)(M\mathcal{G}p)^a(x)\label{nonlocal2}
\end{align}
The second terms on the right hand sides of (\ref{nonlocal1}) and (\ref{nonlocal2}) cancel. Thus from (\ref{cancel1}) and the subsequent algebra, we can conclude that
\beq
[q', T_1+ T_2] = +\frac{i}{e^2}\int f^{kmn}(M^{\dagger -1}\chi)^k(M\chi^\dagger)^m(M^{\dagger -1}\chi)^m + 2 i\int  (M^{\dagger -1}\chi)^m(DM\mathcal{G}p)^m
\eeq
These terms are precisely canceled by
\bsp
\frac{1}{e^2}[q', \int\chi^a\mathcal{D}_J\chi^a] & = [\frac{i}{e^2}\int \chi^{\dagger m}(\mathcal{G}p)^m, \int\chi^a\mathcal{D}_J\chi^a] \\
&=  -\frac{i}{e^2}\int f^{kmn}(M^{\dagger -1}\chi)^k(M\chi^\dagger)^m(M^{\dagger -1}\chi)^m - 2 i\int  (M^{\dagger -1}\chi)^m(DM\mathcal{G}p)^m
\end{split}
\eeq
Finally, it is straightforward to see that
\beq
[q',  \frac{2\pi^2}{e^2c_A^2}\int(\bar \partial J^a \bar \partial J^a)  -\frac{1}{e^2}\int \chi ^{\dagger m} \bar{\mathcal{D}}_{\bar{J}}\chi ^{\dagger m} ] = 0
\eeq
Putting all tis together, we have demonstrated that
\beq
[q', \mathcal{H}] = 0
\eeq
We can thus be assured that (\ref{qkkn}) and its adjoint computed with respect to the measure
 (\ref{measure}), along with (\ref{invh1}), do give a realization of the $\mathcal{N} = 1$ algebra that we started with. Furthermore, note that in the case of the standard presentation of the algebra using the original variables $E, \Psi, A$, etc., the commutator of the supercharge with the Hamiltonian  only vanishes up to the Gauss law generator. In the present case, since we have already eliminated  $E$ using Gauss law, the relevant commutator is identically zero.

It is also worth noting that in the minimally supersymetric case, the number $n$ is $not$ fixed by the constraints of supersymmetry alone. However we shall see in the next section that,
for theories with extended supersymmetry,  it is fixed simply by demanding that the gauge-invariant Hamiltonian commute with the supercharges. However, for the
${\mathcal N} =1$ case,
we have to rely on the independent arguments given earlier in the paper for the volume measure,
as well as consistency with the previous perturbative results \cite{lee-scs};
these imply that $n = 1$ for $\mathcal{N} = 1$.

A nonvanishing coefficient for $S_{wzw}(H)$ in the volume measure implies that the theory will have a mass-gap, with the scale of the massive excitations set by the renormalized level 
number $(k+ N) \, (e^2/4\pi)$. This is also clear from the form of the Hamiltonian (\ref{invh1}). In \cite{witten-index, N=2-vac}, it was pointed out that, for small enough values of the level number, the theory can have spontaneously broken supersymmetry. Since the statement about existence of the mass-gap in the present formalism crucially uses manifest supersymmetry, we cannot comment on the possibility of supersymmetry breaking based solely on the results obtained so far. Presumably a computation of the vacuum wave functional using the present formalism might allow one to study this interesting dynamical question.

 \section{Supercharges and Hamiltonians with extended supersymmetry}
 
We will now consider the extension of the analysis done so far to $\mathcal{N}=2$ and $4 $ supersymmetries.  The terms in the action, for the theories of interest to us, are
 \bsp
S_{YM} &= -\frac{1}{4e^2} \int F^a_{\mu \nu}F^{a\mu \nu} - \frac{1}{2e^2} \int D_\mu \phi ^a_A D^\mu \phi^a_A + \frac{1}{2e^2} \int F^a_AF^a_A \\
& -\frac{i}{2e^2} \int \bar  \psi ^a_I\gamma ^\mu D_\mu \psi ^a_I -\frac{i}{2e^2} \int \bar  \omega ^a\gamma ^\mu D_\mu \omega^a - \frac{i}{2e^2} \int \epsilon_{ABC}\bar\psi  _A^a\psi  _B^b \phi^c_C f^{abc}\\
& +\frac{i}{e^2}\int \bar{\psi }^a_A\omega ^b \phi^a_Af^{abc} - \frac{1}{4e^2}\int f^{abc}f^{amn} \phi^b_B\phi^c_C \phi^m_B\phi^n_C\\
S_{CS} &= -{k \over 4\pi}  \epsilon^{\mu \nu \rho} \int \Tr\,(A_\mu \partial_\nu A_\rho + \frac{2}{3} A_\mu A_\nu A_\rho)\\
&+ \frac{m_k}{2e^2}\int\left(- i\bar\psi ^a _I \psi _I^a+ i\bar \omega^a\omega^a + 2F^a_A\Phi^a_A - \frac{1}{3}f^{abc}\epsilon_{ABC}\phi^a_A\phi^b_B \phi^c_C\right) 
\end{split}
\label{ext1}
\eeq
(As indicated earlier, $m_k = e^2 k /4 \pi$.) The capital Latin indices  take on three  values, which correspond to the manifest $SO(3)$ $R$-symmetry of the theory. The theory has four adjoint Majorana fermionis $\{\psi^a_I, \omega^a\}$. Setting $\omega, \phi_1, \phi_2 , \psi _3$ to  zero  truncates the action to have $\mathcal{N}=2$ supersymmetry. If one further sets $\phi_3, \psi _2$ to zero then we recover the $\mathcal{N}=1$ theory discussed earlier. $F^a_A $ are auxiliary fields, which may be replaced by their saddle point values $F^a_A = - m\, \Phi^a_I$. 

After  absorbing the factor of $\exp\left[ {\frac{k}{2}S_{wzw}(H)}\right]$ in the measure, as we did
 in the $\mathcal{N} = 1$ theory, we obtain the following expression for the $\mathcal{N}=4$ supercharge,
\begin{align}
q'_I &= i\int   \psi^{ a \dagger}_I \frac{\delta }{\delta A^a} + \frac{1}{e^2} \int \psi ^a_I B^a + \epsilon_{IJK}\int \psi ^a_J \left( \Pi^a_{\phi_K} + i {m_k \over e^2} \phi^a_K  \right)
+ \frac{2i}{e^2} \epsilon_{IJK}\int \psi^{a\dagger}_J (\bar D \phi_K)^a\nonumber\\
&\hskip .2in -\int \omega ^a \left( \Pi^a_{\phi_I} - i\frac{m_k}{e^2} \phi^a_I\right)
 - \frac{2i}{e^2}\int \omega ^{a \dagger }(\bar D \phi )^a \label{ext2}\\
&\hskip .2in +  \frac{i}{2e^2}\int f^{abc}\epsilon_{ABI} \phi ^b_A\phi^c_B\omega ^a + \frac{i}{e^2} \int f^{abc} \phi ^b_K \phi^c_I \psi^a_K\nonumber
\end{align}
These give the three $SO(3)$ covariant supercharges; there
is a also a fourth supercharge that commutes with the Hamiltonian.
Our purpose is to use these supercharges to identify the gauge-invariant variables for the 
fermions, extending what we did for the minimally supersymmetric case, and, eventually, the Hamiltonian.
For this, the three charges given above are adequate.
Further, it is tis supercharge which survives upon truncation to lower supersymmetries.
Because of this, we see immediately that, if we use supersymmetry to pick out the choice of
fermionic variables as we did earlier, 
the gauge invariant counterparts of $\psi_I$ are $\lambda _I = \psi_I M^{\dagger  -1}$ and $\lambda ^\dagger _I = \psi^\dagger _I M$, just as in the $\mathcal{N}=1$ case.
As a result, for the case of $\mathcal{N}=2$ supersymmetry (where the labels $I, J, \cdots = 1,2$), 
 the fermions contribute $e^{-2\times c_AS_{wzw}(H) }$ to the volume measure, 
 canceling completely the contribution from the gauge fields. This result implies the masslessness of the $\mathcal{N}=2$, $k=0$ theory.
In the $\mathcal{N}=4$ case, we have an additional contribution of $e^{- c_AS_{wzw}(H) }$ from the third fermion field $\psi_3$. Further, there is a contribution from the remaining fermion field
$\omega$.
Since $\omega $ does not couple to the gauge fields directly in the supercharge, its gauge invariant form has to be deduced from its coupling to the scalar fields.  For this, we concentrate
on two terms in the supercharge given by
\beq
q'_I =  \epsilon_{IJK}\int \psi ^a_J \left( \Pi^a_{\phi_K} + i {m_k \over e^2} \phi^a_K  \right)
 -\int \omega ^a \left( \Pi^a_{\phi_I} - i\frac{m_k}{e^2} \phi^a_I\right)
+\cdots
\label{ext3}
\eeq
Using the already-settled-upon change of variables 
$\psi \rightarrow \chi_I$, this shows that $M^\dagger {\mathbb A}$ is the appropriate gauge-invariant version of
${\mathbb A} =
\left( \Pi^a_{\phi_K} + i (m_k / e^2) \phi^a_K  \right)$. The second term on the right hand side of
(\ref{ext3}), which involves ${\mathbb A}^\dagger$ then tells us that
the gauge-invariant combination for $\omega$ is 
\beq
 \chi_\omega  = M^{-1}\omega \label{ext4}
\eeq
This corresponds to what we referred to as choice I in section 3.
Thus its contribution to the measure is a factor $e^{ c_AS_{wzw}(H) }$, the exponent having the opposite sign to the other three fermions.
The net result is that, as for the ${\mathcal{N} = 2}$ case, the fermion and gauge field
contributions involving $S_{wzw}(H)$ cancel out completely in the
the volume measure for the
 $\mathcal{N}=4$ theory.
 (If $k =0$, this means that the measure simply has $d\mu(H)$ and the fermionic fields, and the theory remains massless  quantum mechanically.)
This is completely consistent with all other indications obtained in the literature using alternate methods, as well as our analysis in section 2.
We will see shortly that this conclusion is  reinforced by the supersymmetry algebra, just as in the minimal case.

Proceeding with this choice of gauge invariant fermionic variables we can write down the the gauge invariant form of the $SO(3)$ covariant  supercharge as
\bsp
q'_I &= \int \chi^{a \dagger}_I (\mathcal{G}p)^a - \frac{2\pi}{e^2c_A}\int \chi^a_I(\bar \partial J)^a + \epsilon_{IJK}\int \chi ^a_J\Pi^a_{\Phi_K} + \epsilon_{IJK}\int \chi ^{a \dagger}_J(H^{-1})^{ab}\bar \partial \Phi^b_K\\
&-\int \chi_\omega ^a(H^{-1})^{ab}\Pi^b_{\Phi_I} - \frac{2i}{e^2}\int \chi^{a \dagger}_\omega \bar \partial \Phi_I^a + \frac{im_k}{e^2}\epsilon_{IJK}\int \chi^a_J\Phi^a_K + \frac{im_k}{e^2}\int \chi_\omega ^a (H^{-1})^{ab}\Phi^b_I \\
&+ \frac{i}{e^2} \int f^{abc} \Phi ^b_K \Phi^c_I \chi^a_K +  \frac{i}{2e^2}\int f^{abc}\epsilon_{ABI} \Phi ^b_A\Phi^c_B(H\chi_\omega )^a
\end{split}
\label{ext5}
\eeq
In this equation, $\Phi_L = M^\dagger \phi_L$ and $\Pi_{\Phi_L} = M^{\dagger}\Pi_{\phi_L}$
are the gauge invariant versions of the scalar fields and their momenta, respectively.
The Gauss law constraint is given by
\beq
(D\frac{\delta }{\delta A} + \bar D\frac{\delta }{\delta \bar A})^a + \frac{1}{e^2} f^{amn}\left(\psi^{ m \dagger}_L\psi^n_L + \omega^{ m \dagger}\omega^n + e ^2 \phi^m_L\frac{\delta}{\delta \phi^n_L}\right) \approx  0
\label{ext6}
\eeq
In terms of action on functions of the gauge-invariant variables we have introduced,
this translates into
 \beq
 \bar p^a = (Hp)^a + \frac{1}{e^2}f^{amn} \left( (H\chi_L^\dagger)^m\chi_L^n + \chi^{m \dagger}_\omega (H\chi_\omega)^n + e^2 \Phi^m_L \frac{\delta}{\delta \Phi^n_L}\right)\label{gauss}
 \eeq
(We may regard the latter form as the requirement of holomorphic invariance of physical wave functionals.)
 
Paralleling the  discussion of the $\mathcal{N} = 1$ theory, our strategy is to take the measure of integration to be of the form (\ref{measure}), with $n$ considered to be arbitrary and then compute the adjoint of the supercharges and enforce the supersymmetry algebra to determine
$n$. (We expect $n$ to vanish from what has already been said, but we do not want to presume this at this stage.)
After the use of Gauss law to eliminate $\bar p$ and the subsequent normal ordering - these manipulations exactly parallel the $\mathcal{N}=1 $ case studied before - we get
\begin{align}
q^{\prime \dagger} _I &=  - i \int \chi_I^a(\bar{\mathcal{G}}Hp)^a - \int\left(1-\frac{n}{2} + \frac{k}{2c_A}\right) \chi^a_I J^a\nonumber\\
&\hskip .2in -\frac{i}{e^2}\int f^{amn}\bar{\mathcal{G}}(x,y)\left((H\chi_L^\dagger)^m\chi_L^n + \chi^{m \dagger}_\omega (H\chi_\omega)^n + e^2 \Phi^m_L {\delta \Phi^n_L}\right)(y)\chi^a_I(x)\nonumber\\
&\hskip .2in +\bigg(- \frac{2\pi}{e^2c_A}\int \chi^a_I(\bar \partial J)^a + \epsilon_{IJK}\int \chi ^a_J\Pi^a_{\Phi_K} + \epsilon_{IJK}\int \chi ^{a \dagger}_J(H^{-1})^{ab}\bar \partial \Phi^b_K\label{ext8}\\
&\hskip .2in -\int \chi_\omega ^a(H^{-1})^{ab}\Pi^b_{\Phi_I} - \frac{2i}{e^2}\int \chi^{a \dagger}_\omega \bar \partial \Phi_I^a + \frac{im_k}{e^2}\epsilon_{IJK}\int \chi^a_J\Phi^a_K + \frac{im_k}{e^2}\int \chi_\omega ^a (H^{-1})^{ab}\Phi^b_I \nonumber\\
&\hskip .2in + \frac{i}{e^2} \int f^{abc} \Phi ^b_K \Phi^c_I \chi^a_K +  \frac{i}{2e^2}\int f^{abc}\epsilon_{ABI} \Phi ^b_A\Phi^c_B(H\chi_\omega )^a\bigg)^\dagger\nonumber
\end{align}
It is also instructive to rewrite this  back in terms of the gauge-covariant original variables. 
The adjoint supercharge then takes the form
\begin{align}
q^{\prime \dagger} _I &= -i\int \psi ^a_I(x)(\bar D ^{-1})^{ab}(x,y) (D\frac{\delta}{\delta A})^b(y) - \frac{ic_A}{\pi}\int \left(1-\frac{n}{2} + \frac{k}{2c_A}\right)\psi_I^a(A-a)^a\nonumber\\
&\hskip .2in -\frac{i}{e^2}\int (\bar D ^{-1})^{ab}(x,y) f^{bmn}\left(\psi^{ m \dagger}_L\psi^n_L + \omega^{ m \dagger}\omega^n + e ^2 \phi^m_L\frac{\delta}{\delta \phi^n_L}\right)(y)\psi^a_I(x)\nonumber\\
&\hskip .2in + \frac{1}{e^2} \int \psi ^{a\dagger}_I B^a + \epsilon_{IJK}\int \psi ^{a\dagger}_J \Pi^a_{\phi_K} - \frac{2i}{e^2} \epsilon_{IJK}\int \psi^{a}_J (D \phi_K)^a\label{ext9}\\
&\hskip .2in -\int \omega ^{a\dagger} \Pi^a_{\phi_I} + \frac{2i}{e^2}\int \omega ^{a  }( D \phi )^a - \frac{im_k}{e^2}\epsilon_{IJK}\int \psi^{a\dagger}_J\phi^a_K - \frac{im_k}{e^2}\int \omega ^{a\dagger} \phi^a_I \nonumber\\
&\hskip .2in -  \frac{i}{2e^2}\int f^{abc}\epsilon_{ABI} \phi ^b_A\phi^c_B\omega ^{a\dagger} - \frac{i}{e^2} \int f^{abc} \phi ^b_K \phi^c_I \psi^{a\dagger}_K\nonumber
\end{align}
We can now construct the Hamiltonian from the anticommutator of the
supercharges given above. It is a straightforward, but lengthy, calculation; the result
is ${\mathcal H} = {\mathcal H}_0 + {\mathcal H}_m$, where, in terms of the gauge-covariant variables,
\begin{align}
{\mathcal H}_0 &= \frac{e^2}{2}\int (\bar D ^{-1})^{ab}(x,y) \left(D\frac{\delta}{\delta A}\right)^b_y \frac{\delta}{\delta A^a(x)}\nonumber\\
&\hskip .2in+\frac{1}{2}\int  (\bar D ^{-1})^{ab}(x,y) f^{bmn}\left(\psi^{ m \dagger}_L\psi^n_L + \omega^{ m \dagger}\omega^n + e ^2 \phi^m_L\frac{\delta}{\delta \phi^n_L}\right)_y \frac{\delta}{\delta A^a(x)}\nonumber\\
&\hskip .2in +\frac{1}{e^2}\int \epsilon_{ABC}f^{amn}\phi^m_A\psi^{a\dagger }_B\psi^n_C +\frac{1}{e^2}\int f^{abc}\left(\phi^c_A\psi^b_A\omega^{a\dagger} + \phi^c_A\omega^a\psi^{b\dagger}_A\right)\label{Hamil1}\\
&\hskip .2in+ \frac{1}{2e^2}\int B^aB^a + \frac{2}{e^2}\int (D\phi_L)^a(\bar D \phi_L)^a + \frac{e^2}{2}\int \Pi_{\phi_L}^a  \Pi_{\phi_L}^a\nonumber\\
&\hskip .2in+\frac{1}{e^2}\int (\psi^a_LD\psi^a_L - \psi^{a \dagger}_L\bar{D}\psi^{a\dagger}_L) +\frac{1}{e^2}\int (\omega^aD\omega^a - \omega^{a \dagger}\bar{D}\omega^{a\dagger})
\nonumber\\
&\hskip .2in+\frac{1}{4}f^{amn}f^{apq}\phi^m_M\phi^n_N\phi^p_M\phi^q_N\nonumber
\end{align}
\bsp
{\mathcal H}_m &= \frac{e^2(k + (2-n)c_A)}{4\pi}\int(A-a)^a\frac{\delta}{\delta A^a} + \frac{e^2(k + (2-n)c_A)}{4\pi e^2}\int \psi^{a\dagger}_L\psi^a_L\\
&\hskip .2in -\frac{k}{4\pi}\int \omega^{a\dagger}\omega ^a + \frac{1}{2}\left[\frac{k}{4\pi}\right]^2e^2\int \phi^a_I\phi^a_I
\end{split}
\label{Hamil2}
\eeq
The second term of ${\mathcal H}$, which is the mass term
${\mathcal H}_m$, we see that the masses of the gauge field and the $SO(3)$ fermions get a shift proportional to $(2-n)c_A$ while the scalars and the fourth fermion do not. Obviously this is the result of the fact that only the $SO(3)$ fermions couple to the electric field in the supercharge, and are hence affected by the singular contributions proportional to $\bar D^{-1}(x,x)$ brought about by the use of the Gauss law and normal ordering, see the discussion following
(\ref{bos-mass}).
 For the theory to be supersymmetric, one must necessarily have equal
 masses for these degrees of freedom; this is obtained
 only when $n=2$, which makes the mass term $\mathcal{H}_m$ vanish when $k = 0$.
We can also verify more explicitly, with another lengthy calculation, that 
\beq
[q_I, H] = 0 \hspace{.3cm}\Longrightarrow \hspace{.3cm} n=2  \hspace{.2cm}\mbox{for}  \hspace{.2cm} \mathcal{N}\geq 2\label{ext10}
\eeq
Thus in the case of extended supersymmetry, the requirement of supersymmetry invariance forces the measure to be exactly what we had presented earlier from anomaly considerations, leading to a gapless spectrum for the gauge theories in the $k = 0$ limit.

For completeness, we also give here
the formulae for ${\mathcal H}_0$ and ${\mathcal H}_m$ in terms of the gauge invariant variables.
\begin{align}
{\mathcal H}_0 &= \frac{e^2c_A}{2\pi}\int \Omega ^{ab}(x,y)\frac{\d}{\d J^a(x)}\frac{\d}{\d J^b(y)}\nonumber\\
&\hskip .2in +\frac{ic_A}{2\pi}\int \bar{\mathcal{G}}(x,y)f^{abc}\left((H\chi^{b\dagger}_L)\chi^c_L + \chi_\omega ^{b \dagger}(H\chi_\omega)^c + \Phi^b_L\frac{\d}{\d \Phi^c_L}\right)(y)\frac{\d}{\d J^a(x)}
\nonumber\\
&\hskip .2in -\frac{1}{e^2}\int f^{abc}\epsilon_{ABC}\Phi^a_A(H\chi^\dagger)^b_B\chi^c_C - \frac{1}{e^2}\int f^{abc}(\Phi^a_A\chi^b_A\chi^\dagger _\omega +  \mbox{h.c})\label{invn=40}\\
&\hskip .2in +\frac{2\pi^2}{e^2c_A^2} \int \bar{\partial}J^a \bar{\partial}J^a + \frac{2}{e^2}\int \bar{\partial}\Phi^a_A(\mathcal{D}_J \Phi_L)^a + \frac{e^2}{2}\int \Pi^a_{\Phi_L} \Pi^a_{\Phi_L}\nonumber\\
&\hskip .2in + \frac{1}{e^2}\int (\chi^a_L(\mathcal{D}_J\chi_L)^a - \chi^{a\dagger}_L(\bar{\mathcal{D}}_{\bar{J}}\chi_L)^{a\dagger}) + \frac{1}{e^2}\int(\chi_\omega ^a\partial \chi^a_\omega - \chi_\omega ^{a\dagger}\bar{\partial} \chi^{a\dagger}_\omega)\nonumber\\
&\hskip .2in +\frac{1}{4e^2}\int f^{lpq}f^{kbc}\Phi^p_K\Phi^q_I\Phi^{\dagger b}_K\Phi^{\dagger c}_IH^{lk}
\nonumber\\
{\mathcal H}_m &=  \frac{e^2(k + (2-n)c_A)}{4\pi}\int \left[ J^a\frac{\d}{\d J^a} + \frac{1}{e^2} \chi^{a \dagger }_L(H^{-1})^{ab}\chi^b_L\right]\nonumber\\
&\hskip .2in -\frac{k}{4\pi}\int \chi^{a\dagger}_\omega H^{ab}\chi^b_\omega + \frac{1}{2}\left[\frac{k}{4\pi}\right]^2e^2\int \Phi^a_I\Phi^a_I
\label{invn=4m}
\end{align}

\section{Linearization with $\mathcal{N} = 1,2,4$ - Spectrum and the Algebra}

As in the case of the purely gluonic theory, one can consistently linearlize the 
Hamiltonian and the supercharges in the supersymmetric theories. This linearization
 will yield a purely algebraic justification for the measure.
Following \cite{KKN-long}, one can define $H = e^{t^a\phi^a }$ ( $\phi  \sim (\theta + \bar \theta)$ defined in (\ref{variationwzw})) and expand the Hamiltonian as well as the supercharges to quadratic order in the $\phi^a$'s. This produces an abelian-dualized theory which can be put in the familiar form involving four scalar fields for the $\mathcal{N} = 4$ case upon reabsorbing the measure factor $e^{[k + (2-n)c_A] S_{wzw}(H)}$ in the wave functionals and defining the scalar field as
\beq
\Phi_H = \sqrt{k {\bar k} / e^2}~ \,\phi.\label{lin1}
\eeq
(Here $k$, ${\bar k}$ denote the Fourier transforms of $\del$ and $\bdel$, respectively.)
The dualized Hamiltonian for the theory  becomes that of a free theory of four scalars $(\Phi_H, \Phi_I)$, with masses 
given by 
\beq
m_n = (k + (2-n)c_A) {e^2\over 4\pi}
\label{lin2}
\eeq
The fermionic part of the dualized theory is nothing but the naive truncation of the corresponding parts of (\ref{invn=40}, \ref{invn=4m}) to quadratic levels. It is understood, as before, that $n = 1 $ for $\mathcal{N}=1$ and $n=2$ for $\mathcal{N}\geq 2$.

At this point it is very instructive to look at the supersymmetry algebra for the linearized theory. The action corresponding to the free and linearized theory is
\bsp
S_{lin} &= \int \frac{1}{2}\Phi^a_H (\del^\mu \partial_\mu - m_n^2)\Phi^a_H + \frac{1}{2}\Phi^a_I (\del^\mu \partial_\mu - m_n^2)\Phi^a_I\\
&-\frac{i}{2}\int \bar\l^a_I(\del^\mu \gamma_\mu -m_n)\l^a_I + \bar\omega^a(p^\mu \gamma_\mu +m_n)\omega^a
\end{split}
\label{lin3}
\eeq
with the previously mentioned constraints on $n$.
The supersymmetry transformation laws for $S_{lin}$ deduced from the linearization of $q_I$  are
\bsp
\delta_\e \Phi^a_H &= \frac{i}{2}\bar{\l}^a_I\e_I\\
\delta_\e \Phi^a_I &= \frac{i}{2}(\e_{IJK}\bar{\l}^a_J\e_K + \bar\omega^a\e_I)\\
\delta \l^a_I &= \frac{1}{2}(\gamma^\mu\del_\mu + m_n) (\Phi^a_H\e_I - \epsilon_{IJK}\Phi_J^a\e_K)\\
\delta\omega^a &= \frac{1}{2}(\gamma^\mu\del_\mu - m_n)\Phi_I \e_I
\end{split}
\eeq
The closure of the algebra on the scalars gives
\bsp
[\delta_\b, \d_\e]\Phi^a_H  & = \frac{i}{2}\left(\bar \e_I\gamma^\mu\b_I\right) \del_\mu \Phi^a_H + \frac{im_n}{2}\bar \e_I\b_K\e_{IJK}\Phi^a_J\\
[\delta_\b, \d_\e]\Phi^a_I  & = \frac{i}{2}\left(\bar \e_I\gamma^\mu\b_I\right) \del_\mu \Phi^a_I + \frac{im_n}{2}\bar \e_K\b_J\e_{IJK}\Phi^a_H
\end{split}\label{defalg}
\eeq
Thus we see that the algebra, instead of simply closing on the momentum generators, involves a non-central extension generated by three extra $U(1)$ generators that mix the dual gauge field 
$\Phi_H$ with the three $SO(3)$ covariant scalars. These extra $U(1)$ symmetries are {\it not} visible in the Hamiltonian of the theory, however, they can be interpreted as symmetries of the on-shell $S$-matrices of the theories under consideration.  
In the limit of $k = 0$ the three `hidden' $U(1)$ generators couple to the manifest $SO(3)$ symmetries to generate an $SO(4)$ symmetric $S$-matrix for the $\mathcal{N} = 4$ theory 
to {\it all} orders in perturbation theory.  As a matter of fact, the $SO(\mathcal{N})$ invariance of the $S$-matrices of the $\mathcal{N} = 2,4,8$ theories to all perturbative orders was explicitly shown in \cite{so(8)}. 

For the purposes of our present discussion, we see that the appearance of the mass on the 
right hand side of (\ref{defalg}) implies that the renormalized level-number $(k + (2-n)c_A)$ plays the role of a structure constant. 
However the algebra (\ref{defalg}) must be satisfied by the linearization of the dualized theory at every perturbative order, including the tree level theory. As shown earlier in this paper, the renormalization of the level number $(k \rightarrow (k + (2-n)c_A))$ arises from a Jacobian and hence the term proportional to $c_A$ in the renormalized level number is to be regarded 
as  $\mathcal{O}(\hbar)$. It can be readily seen that applying a dualization prescription $A = -\partial \theta$ to the tree level $\mathcal{N} =4 $ Hamiltonian, obtained using the canonical quantization procedure reviewed in (\ref{canonical}), produces  (\ref{defalg}) as the symmetry algebra of the quadratic part of the theory, but with the unrenormalized level number.
Since structure constants cannot undergo quantum corrections, the only way to reconcile these
statements, namely, the consistent appearance of (\ref{defalg}) as the symmetry algebra of the linearized theory to all orders in perturbation theory and the nonrenormalization of a structure constant is for $n$ to be $2$ for $\mathcal{N}\geq 2 $ supersymmetry. 
We thus see that the results obtained earlier on the effect of the fermions on the measure can also justified on purely algebraic grounds. 

Finally, we note that
that the appearance of the noncentral extension in the the algebra  (\ref{defalg}) is tied in
with the parity violating nature of the fermion mass-terms. If one began with the free,
massive $\mathcal{N}=1$ chiral multiplet in four dimensions, its dimensional reduction (in our conventions) would produce a parity conserving  $\mathcal{N} = 2$ mass term 
$\sim \int (\l^\dagger_1 \l_1 -  \l^\dagger_2\l_2)$ in three dimensions. The closure of the SUSY algebra with such masses would not result in the extension we have above. It is precisely the parity violating nature of the $\int \l^\dagger_I\l_I $ mass term, which, in turn, is dictated by the  parity violating nature of the Chern-Simons term that leads to the massive on-shell algebra.
\bigskip\\
{\it \underline {Acknowledgements}:} This research was supported by National Science Foundation grant PHY-0855515 and by PSC-CUNY grants.
\section*{\normalsize APPENDIX}
\def\theequation{A\arabic{equation}}
\setcounter{equation}{0}

Here we collect some useful formulae and commutation relations that have been used throughout the paper. An exhaustive list of the commutation relations between various operators used in the gauge invariant framework can be found in\cite{KKN-long}
\bsp
&[p^a(x), H(y)] = H(y)(-it^a) \d^{(2)}(x-y), \hspace{.3cm} [p^a(x), M(y)] = M(y)(-it^a) \d^{(2)}(x-y)\\
&[p^a(x), H^{bc}(y)] = f^{acd}H^{bd}(y) \d^{(2)}(x-y),  \hspace{.3cm} [p^s(x), J^b(y)] = -i\frac{c_A}{\pi}H^{bs}(y)\, \partial_y\d^{(2)}(y-x)
\end{split}\label{pcomm}
\eeq
The definition of $J$ used throughout the paper is 
\beq
J^a = \frac{2\,c_A}{\pi}\Tr(t^a\partial H H^{-1}) =  \frac{ic_A}{\pi}(M^\dagger)^{ab}(A-a)^b
\eeq
This is related to the magnetic field as
\beq
B^a = F^a_{12} = -2\,i\,(\bar D A- \partial \bar A)^a = -\frac{2\pi}{c_A}(M^{\dagger -1})^{ab}(\bar \partial J)^b.
\eeq
The regulated Green function  for the $\bar D$ operator is
\bsp
(\bar D ^{-1})^{ab}(x,y) &= (M^{\dagger -1})^{ac}(x)\, \bar{\mathcal{G}}^{cd}(x,y)\, (M^{\dagger})^{db}(y)\\
\bar{\G} _{ab} (\vx,\vy)  &= {1\over \pi (x-y)}   \Bigl[ \d _{ab} - e^{-|\vx-\vy|^2/\e} \bigl(
H(x,\by) H^{-1} (y, \by) \bigr) _{ab}\Bigr]
\end{split}
\eeq
where $\epsilon$ is a regulator parameter, to be taken to zero at the end.
The regulated value of $\bD^{-1}$ at coincident points is
\beq
(\bar D ^{-1})^{ab}(x,x) = -\frac{1}{\pi}f^{abc}(A-a)^c(x) = +\frac{i}{c_A}f^{abc}(M^{\dagger -1})^{cl}(x)J^l(x).
\eeq
In establishing the supersymmetry invariance of the gauge-invariant forms of the Hamiltonians,
we have repeatedly used the identity
\beq
f^{gak}(M)^{ck}(M^{ -1})^{gm} = -f^{mcl} M^{la}
\eeq
A similar identity holds for
$M^\dagger$.
We have also used the fact that $M^T = M^{-1}$ for the adjoint version of $M$.
%
%


\begin{thebibliography}{99} 

 \bibitem{KKN-long}
  D.~Karabali and V.~P.~Nair,
  Nucl.\ Phys.\  B {\bf 464}, 135 (1996)
  [arXiv:hep-th/9510157];
  Phys.\ Lett.\ B {\bf 379}, 141 (1996)
  [hep-th/9602155];
   D.~Karabali, C.~j.~Kim and V.~P.~Nair,
  Nucl.\ Phys.\  B {\bf 524}, 661 (1998)
  [arXiv:hep-th/9705087].
  
  \bibitem{qgp} 
  D.~J.~Gross, R.~D.~Pisarski and L.~G.~Yaffe,
  Rev.\ Mod.\ Phys.\  {\bf 53}, 43 (1981);
   T.~Appelquist and R.~D.~Pisarski,
  Phys.\ Rev.\ D {\bf 23}, 2305 (1981);
    P.~B.~Arnold and L.~G.~Yaffe,
  Phys.\ Rev.\ D {\bf 52}, 7208 (1995)
  [hep-ph/9508280];
   G.~Alexanian and V.~P.~Nair,
  Phys.\ Lett.\  B {\bf 352}, 435 (1995)
  [arXiv:hep-ph/9504256];
  V.~P.~Nair,
  Phys.\ Lett.\  B {\bf 352}, 117 (1995)
  [arXiv:hep-th/9406073].
  
  \bibitem{blg}
  J.~Bagger, N.~Lambert,
  Phys.\ Rev.\  {\bf D75}, 045020 (2007).
  [hep-th/0611108];
  Phys.\ Rev.\  {\bf D77}, 065008 (2008).
  [arXiv:0711.0955 [hep-th]];
  A.~Gustavsson,
  Nucl.\ Phys.\  {\bf B811}, 66-76 (2009).
  [arXiv:0709.1260 [hep-th]];
  J.~Bagger, N.~Lambert,
  JHEP {\bf 0802}, 105 (2008).
  [arXiv:0712.3738 [hep-th]].
  
\bibitem{abjm}
  O.~Aharony, O.~Bergman, D.~L.~Jafferis, J.~Maldacena,
  JHEP {\bf 0810}, 091 (2008).
  [arXiv:0806.1218 [hep-th]].
  
 \bibitem{hightc} 
  C.~P.~Herzog, P.~Kovtun, S.~Sachdev and D.~T.~Son,
  Phys.\ Rev.\ D {\bf 75}, 085020 (2007)
  [hep-th/0701036].
  

  \bibitem{KKN-string}
  D.~Karabali, C.~j.~Kim and V.~P.~Nair,
  Phys.\ Lett.\  B {\bf 434}, 103 (1998)
  [arXiv:hep-th/9804132];
   D.~Karabali, V.~P.~Nair and A.~Yelnikov,
  Nucl.\ Phys.\ B {\bf 824}, 387 (2010)
  [arXiv:0906.0783 [hep-th]].

\bibitem{scalar} 
  A.~Agarwal, D.~Karabali and V.~P.~Nair,
  Nucl.\ Phys.\ B {\bf 790}, 216 (2008)
  [arXiv:0705.0394 [hep-th]].

\bibitem{sphere} 
  A.~Agarwal and V.~P.~Nair,
  Nucl.\ Phys.\ B {\bf 816}, 117 (2009)
  [arXiv:0807.2131 [hep-th]].
  
    \bibitem{glueballs} 
  R.~G.~Leigh, D.~Minic and A.~Yelnikov,
  Phys.\ Rev.\ D {\bf 76}, 065018 (2007)
  [hep-th/0604060].
  
    \bibitem{KKN-CS} 
  D.~Karabali, C.~-j.~Kim and V.~P.~Nair,
  Nucl.\ Phys.\ B {\bf 566}, 331 (2000)
  [hep-th/9907078].
  
    \bibitem{DJT} 
  S.~Deser, R.~Jackiw and S.~Templeton,
  Phys.\ Rev.\ Lett.\  {\bf 48}, 975 (1982),
  S.~Deser, R.~Jackiw and S.~Templeton,
  Annals Phys.\  {\bf 140}, 372 (1982)
  
  \bibitem{robustness} 
  D.~Karabali and V.~P.~Nair,
  Phys.\ Rev.\ D {\bf 77}, 025014 (2008)
  [arXiv:0705.2898 [hep-th]].

 \bibitem{Z} 
  A.~Kapustin, B.~Willett and I.~Yaakov,
  JHEP {\bf 1003}, 089 (2010)
  [arXiv:0909.4559 [hep-th]],
  A.~Kapustin, B.~Willett and I.~Yaakov,
  JHEP {\bf 1010}, 013 (2010)
  [arXiv:1003.5694 [hep-th]],
  D.~L.~Jafferis,
  arXiv:1012.3210 [hep-th],
   S.~Cheon, H.~Kim and N.~Kim,
  JHEP {\bf 1105}, 134 (2011)
  [arXiv:1102.5565 [hep-th]],
    N.~Hama, K.~Hosomichi and S.~Lee,
  JHEP {\bf 1103}, 127 (2011)
  [arXiv:1012.3512 [hep-th]],
    S.~Kim,
  Nucl.\ Phys.\ B {\bf 821}, 241 (2009)
  [arXiv:0903.4172 [hep-th]],
   N.~Drukker, M.~Marino and P.~Putrov,
  Commun.\ Math.\ Phys.\  {\bf 306}, 511 (2011)
  [arXiv:1007.3837 [hep-th]].
  
    \bibitem{so(8)}
  A.~Agarwal, D.~Young,
  JHEP {\bf 1105}, 100 (2011).
  [arXiv:1103.0786 [hep-th]].
 
 \bibitem{scs-s}
  A.~Agarwal, N.~Beisert, T.~McLoughlin,
  JHEP {\bf 0906}, 045 (2009).
  [arXiv:0812.3367 [hep-th]].
  S.~Lee,
  Phys.\ Rev.\ Lett.\  {\bf 105}, 151603 (2010).
  [arXiv:1007.4772 [hep-th]].
   D.~Gang, Y.~-t.~Huang, E.~Koh, S.~Lee, A.~E.~Lipstein,
  JHEP {\bf 1103}, 116 (2011).
  [arXiv:1012.5032 [hep-th]].
  T.~Bargheer, F.~Loebbert, C.~Meneghelli,
  Phys.\ Rev.\  {\bf D82}, 045016 (2010).
  [arXiv:1003.6120 [hep-th]].
Y.~t.~Huang and A.~E.~Lipstein,
  JHEP {\bf 1010}, 007 (2010)
  [arXiv:1004.4735 [hep-th]].
  A.~E.~Lipstein,
  arXiv:1105.3231 [hep-th],
Y.~t.~Huang and A.~E.~Lipstein,
  JHEP {\bf 1011} (2010) 076
  [arXiv:1008.0041 [hep-th]].
  W.~-M.~Chen, Y.~-t.~Huang,
  [arXiv:1107.2710 [hep-th]],
   M.~S.~Bianchi, M.~Leoni, A.~Mauri, S.~Penati, A.~Santambrogio,
  [arXiv:1107.3139 [hep-th]].
  
\bibitem{lm} 
  H.~Lin and J.~M.~Maldacena,
  Phys.\ Rev.\ D {\bf 74}, 084014 (2006)
  [hep-th/0509235].

  \bibitem{spectrum} 
  N.~Gromov and P.~Vieira,
  JHEP {\bf 0901}, 016 (2009)
  [arXiv:0807.0777 [hep-th]];
  JHEP {\bf 0902}, 040 (2009)
  [arXiv:0807.0437 [hep-th]],
   A.~Agarwal and D.~Young,
  Phys.\ Rev.\ D {\bf 82}, 045024 (2010)
  [arXiv:1003.5547 [hep-th]]. 
 
 \bibitem{wilson-loops} 
  N.~Drukker, J.~Plefka and D.~Young,
  JHEP {\bf 0811}, 019 (2008)
  [arXiv:0809.2787 [hep-th]];
 A.~Agarwal and D.~Young,
  JHEP {\bf 0906}, 063 (2009)
  [arXiv:0904.0455 [hep-th]]. 
  
\bibitem{GK} 
  K.~Gawedzki and A.~Kupiainen,
  Phys.\ Lett.\ B {\bf 215}, 119 (1988);
  Nucl.\ Phys.\ B {\bf 320}, 625 (1989).

 \bibitem{Bos-Nair} 
  M.~Bos and V.~P.~Nair,
  Int.\ J.\ Mod.\ Phys.\ A {\bf 5}, 959 (1990).

  
\bibitem{pisarski-rao} 
  R.~D.~Pisarski and S.~Rao,
  Phys.\ Rev.\ D {\bf 32}, 2081 (1985).
  
    \bibitem{witten-top} 
  E.~Witten,
  Commun.\ Math.\ Phys.\  {\bf 117}, 353 (1988).
  
  \bibitem{lee-scs} 
  H.~-C.~Kao, K.~-M.~Lee and T.~Lee,
  Phys.\ Lett.\ B {\bf 373}, 94 (1996)
  [hep-th/9506170].
  

  
   \bibitem{witten-index} 
  E.~Witten,
  In *Shifman, M.A. (ed.): The many faces of the superworld* 156-184
  [hep-th/9903005].
  
  \bibitem{n=1-lattice} 
  J.~W.~Elliott and G.~D.~Moore,
  JHEP {\bf 0711}, 067 (2007)
  [arXiv:0708.3214 [hep-lat]].

 \bibitem{N=2-brane} 
  J.~Gomis and J.~G.~Russo,
  JHEP {\bf 0110}, 028 (2001)
  [hep-th/0109177].  
  
 \bibitem{unsal-bion} 
  M.~Unsal,
  Phys.\ Rev.\ D {\bf 80}, 065001 (2009)
  [arXiv:0709.3269 [hep-th]]. 
  

 \bibitem{seiberg-witten} 
  N.~Seiberg and E.~Witten,
  In *Saclay 1996, The mathematical beauty of physics* 333-366
  [hep-th/9607163]. 
  
  
 \bibitem{feynman}
  R.~P.~Feynman,
  Nucl.\ Phys.\ B {\bf 188}, 479 (1981).
  
\bibitem{config1} I.M. Singer, Physica Scripta {\bf 24} (1980) 817;
I.M. Singer,  \CMP~ {\bf 60} (1978) 7.

\bibitem{config2}
P.K. Mitter and C.M. Viallet, \CMP~ {\bf 79} (1981) 457;
Phys. Lett. {\bf 85B} (1979) 246; M. Asorey and P.K. Mitter, \CMP~
{\bf 80} (1981) 43; O. Babelon and C.M. Viallet, \CMP~{\bf 81} (1981) 515;
Phys. Lett. {\bf 103B} (1981) 45; P. Orland, arXiv:hep-th/9607134;
\PR~{\bf D70}, 045014 (2004).


  \bibitem{N=2-vac} 
  I.~Affleck, J.~A.~Harvey and E.~Witten,
  Nucl.\ Phys.\ B {\bf 206}, 413 (1982),
    J.~de Boer, K.~Hori and Y.~Oz,
  Nucl.\ Phys.\ B {\bf 500}, 163 (1997)
  [hep-th/9703100],
   O.~Aharony, A.~Hanany, K.~A.~Intriligator, N.~Seiberg and M.~J.~Strassler,
  Nucl.\ Phys.\ B {\bf 499}, 67 (1997)
  [hep-th/9703110].
  
 \end{thebibliography}
\end{document}